\documentclass[pra,twocolumn,amsmath,amssymb,superscriptaddress]{revtex4-1}

\usepackage{epsfig,amsmath}
\usepackage{subfigure}
\usepackage{graphicx}
\usepackage{dcolumn}
\usepackage{stmaryrd}
\usepackage{mathrsfs}
\usepackage{pifont}
\usepackage{amsthm}
\usepackage{amssymb}
\usepackage{bm}
\usepackage{latexsym}
\usepackage[colorlinks=true,linkcolor=blue,citecolor=blue]{hyperref}
\usepackage{color}
\usepackage{epstopdf}

\begin{document}

\title{Fast and stable charging via a shortcut to adiabaticity}

\author{Hanyuan Hu}
\affiliation{Department of Physics, Zhejiang University, Hangzhou 310027, Zhejiang, China}

\author{Shi-fan Qi}
\affiliation{Department of Physics, Zhejiang University, Hangzhou 310027, Zhejiang, China}

\author{Jun Jing}
\email{Email address: jingjun@zju.edu.cn}
\affiliation{Department of Physics, Zhejiang University, Hangzhou 310027, Zhejiang, China}

\date{\today}

\begin{abstract}
Quantum battery is an emerging subject in the field of quantum thermodynamics, which is applied to charge, store and dispatch energy in quantum systems. In this work, we propose a fast and stable charging protocol based on the adiabatic evolution for the dark state of a three-level quantum battery. It combines the conventional stimulated Raman adiabatic passage (STIRAP) and the quantum transitionless driving technique. The charging process can be accelerated up to nearly one order in magnitude even under constraint of the strength of counter-diabatic driving. To perform the charging protocol in the Rydberg atomic system as a typical platform for STIRAP, the prerequisite driving pulses are modified to avoid the constraint of the forbidden transition. Moreover, our protocol is found to be more robust against the environmental dissipation and dephasing than the conventional STIRAP.
\end{abstract}

\maketitle

\section{Introduction}

Following the endeavour to reformulate the classical thermodynamics with respect to both fundamental notions and applications into the quantum language, quantum thermodynamics has received an increasing attention with a vast number of achievements made in the last decade~\cite{thermo1,thermo2,thermo3,thermo4}. To generalize the notion of the maximal capacity to exchange energy between an active state and a passive state of an interested system, Alicki and Fannes~\cite{bat1pioneer} pioneered a quantum device named quantum battery that can store and distribute energy as its counterpart in the classical world. In exploiting its essentially higher operational power, more amount of extractable energy, and other potential advantages, many careful and intensive investigations~\cite{bat1pioneer,bat2quancell,bat3,bat4chargemediate,bat5spinchain,bat6,bat7opensys,bat8chargemediate,
bat9mantbody,bat10ultrafast,bat11fluc,bat12fast,bat13stable,bat14Envimediate,bat15fluc,bat16dark,bat17entangle,
bat18solidstate,bat19extractable,bat20harmcharging,bat21cascade} have been recently carried out around this newly-developed subject.

\begin{figure}[htbp]
\centering
\subfigure[Two-level battery]{
\includegraphics[width=0.35\linewidth]{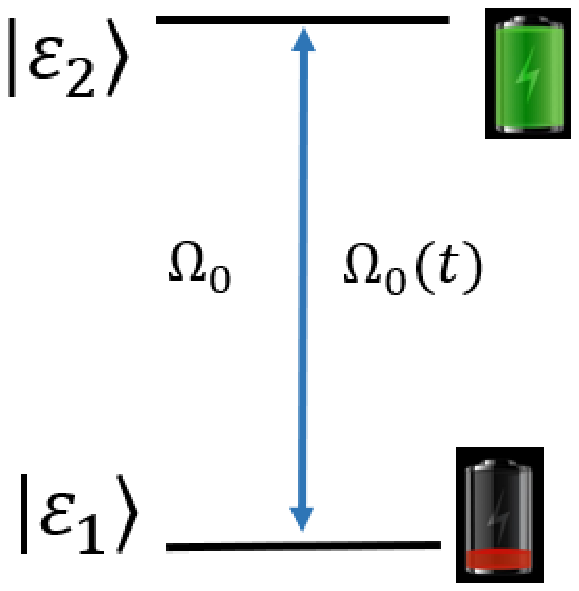}
}
\subfigure[Unstable charging]{
\includegraphics[width=0.55\linewidth]{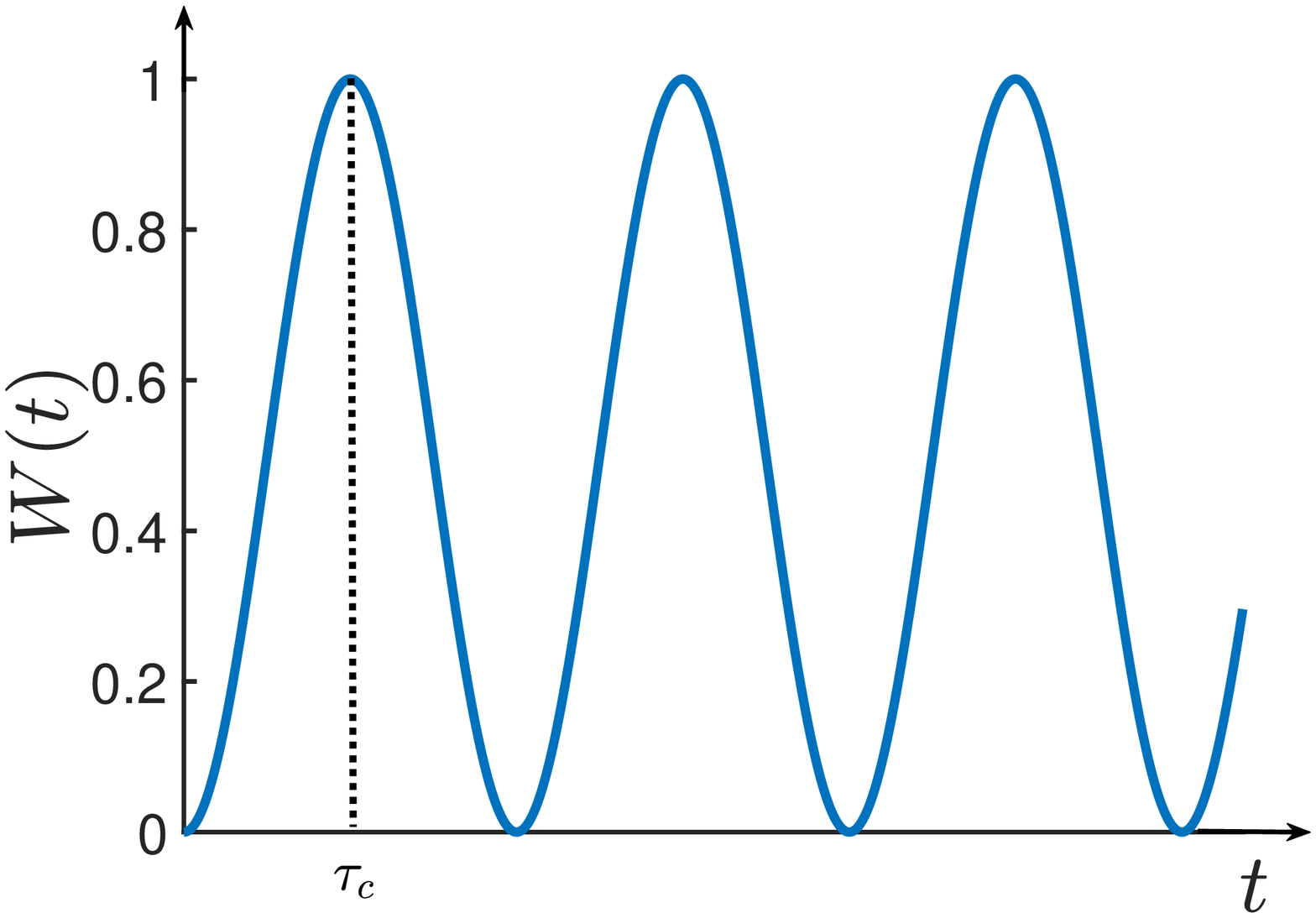}
}
\quad
\subfigure[Three-level battery]{
\includegraphics[width=0.35\linewidth]{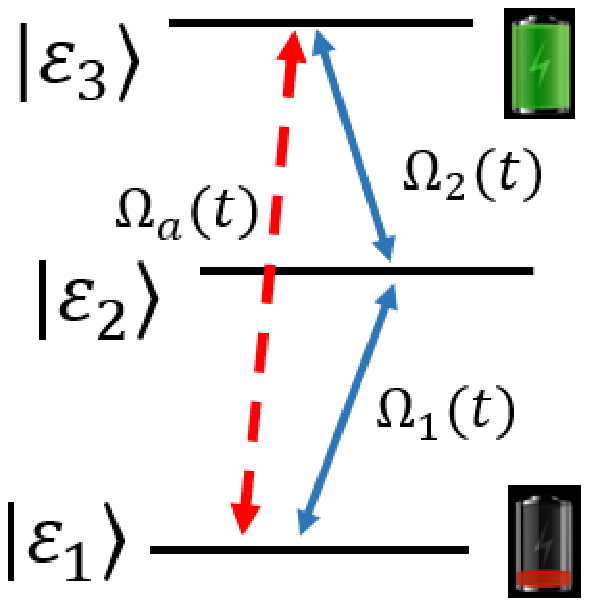}
}
\subfigure[Stable charging]{
\includegraphics[width=0.55\linewidth]{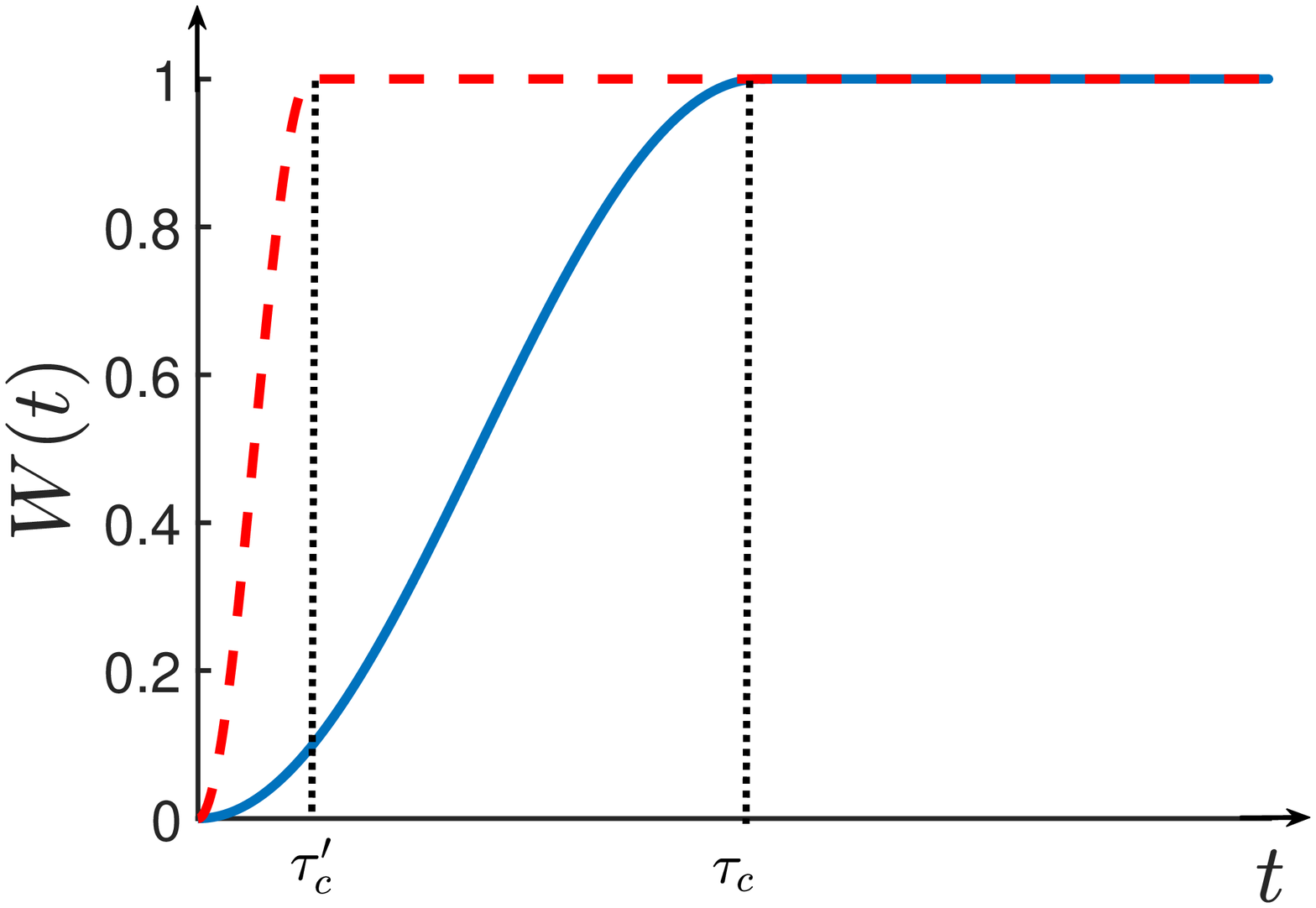}
}
\caption{(Color online) Unstable and stable quantum batteries: (a) A two-level quantum battery coupled to the charger field with time-independent or time-dependent amplitude. (b) Unstable charging process of the two-level quantum battery showing oscillations after the fully-charging time $\tau_c$ in the absence of environments. (c) A cascade three-level quantum battery with two allowed transitions (blue) and an assistant driving for STA (red). (d) A stable charging process of the three-level quantum battery implemented by adiabatic protocols, where the blue-solid line and the red-dashed line represent the STIRAP and our STA protocols, respectively. }\label{fig:demo}
\end{figure}

Quantum batteries were originally established either on a single two-level system [see Fig.~\ref{fig:demo}(a)] (qubit)~\cite{bat4chargemediate,bat8chargemediate,bat21cascade}, or on an array of two-level systems~\cite{bat5spinchain,bat6,bat9mantbody,bat10ultrafast,bat16dark,bat17entangle,bat20harmcharging}. They are charged by external energy sources that are single-frequency optical fields in most cases, and then are discharged in a completely invertible way. Many protocols were applied to implement a fully-charging process, such as the Rabi oscillation of the two-level system stimulated by the static charging field~\cite{bat5spinchain,bat6,bat10ultrafast,bat12fast,bat18solidstate}, the charger-mediated charging due to the interaction between the system and charger qubits~\cite{bat4chargemediate,bat8chargemediate,bat21cascade} and the environment-mediated charging~\cite{bat14Envimediate,bat16dark}, to name a few. Time-dependent~\cite{bat9mantbody,bat20harmcharging} rather than time-independent~\cite{bat5spinchain,bat6,bat10ultrafast,bat12fast,bat18solidstate} parametric laser is exploited to attain a capacity to store more energy in an array of two-level systems. All of the charging protocols mentioned above are yet unstable. As shown in Fig.~\ref{fig:demo}(b), the energy will go back and forth from the battery to the charger if the external field is still on after the charging time $\tau_c$. Thus both the charging and discharging protocols for the two-level systems demand a precise control over the interaction time, a {\em priori} parameter determined by the coupling strength between the battery and the charger as well as their detuning. As a result, the charging performance depends heavily on the ability to decouple the quantum battery from its charger.

To mitigate the unstable charging that might be intrinsic in two-level system, a cascade-type three-level system [see Fig.~\ref{fig:demo}(c)] is introduced by Ref.~\cite{bat13stable} to establish a quantum battery. Soon it was followed by another work~\cite{bat22loop} using a loop-type three-level system. Rather than the Rabi oscillation charging protocol in the two-level system, the authors apply a stimulated Raman adiabatic passage (STIRAP) to facilitate a stable charging protocol. Based on the adiabatic passage of the system dark-state, the charging dynamics is demonstrated by the blue-solid line in Fig.~\ref{fig:demo}(d). The STIRAP protocol turns the full-charging time $\tau_c$ to a {\em posteriori} parameter and thus avoids the infidelity induced by the imprecise control over the interaction time between the battery and the charging fields. But due to the adiabatic condition, this charging protocol is much slower than the Rabi-oscillation based protocol of two-level systems, then and it is not efficient enough for a quantum battery. Meanwhile, during the long-time evolution, the battery is under the influence of quantum decoherence effect, such as energy dissipation and phase decoherence. In this work, we propose an improved STIRAP charging scheme with a much faster charging speed by implanting a technique of shortcut to adiabaticity (STA). Shortcut to adiabaticity~\cite{STA1torron,STA2CMA} in recent years constitutes a toolbox for a fast and robust control method widely applied in quantum optics and quantum information processing~\cite{STA3connect,STA4fastforward,STA5fastforwad,STA6statetransfer,STA7invariant,STA8invariat,STA9complete,
STA10invariant,STA11Berryphase}. Roughly it can be categorized into the counter-diabatic (CD) driving~\cite{STA11adiabatic,STA12assisted,STA13TL,STA14,STA15CD} (also named quantum transitionless driving~\cite{STA16transitionless}), the fast-forward scaling~\cite{STA4fastforward,STA5fastforwad,STA17fastward,STA18fast,STA19acceleration,STA20shortcuts} and the invariant-based inverse engineering~\cite{STA7invariant,STA8invariat,STA10invariant,STA21invariant,STA22fast} with their variants. We will focus in the following on the counter-diabatic driving method. In particular, a counter-diabatic term
\begin{eqnarray}\nonumber
H_{\rm CD}&=&i\sum_{j=1}^n\big[|\partial_t\lambda_j(t)\rangle\langle\lambda_j(t)|\\ \label{eq:HCD} &-&\langle\lambda_j(t)|\partial_t\lambda_j(t)\rangle|\lambda_j(t)\rangle\langle\lambda_j(t)|\big]
\end{eqnarray}
would be added to the reference time-dependent Hamiltonian $H(t)=\sum_{j=1}^n\lambda_j(t)|\lambda_j(t)\rangle\langle\lambda_j(t)|$, where $\lambda_j(t)$ and $|\lambda_j(t)\rangle$ are the $j$th instantaneous eigenvalue and eigenstate of $H(t)$, respectively. Aided by the CD term, the dynamics of quantum state governed by the full Hamiltonian $H_{\rm STA}=H+H_{\rm CD}$ will evolve exactly along the adiabatic path~\cite{STA1torron,STA2CMA} with a much faster speed than that of the conventional STIRAP [see the red-dashed line in Fig.~\ref{fig:demo}(d)]. Moreover, CD driving can promote the system robustness against the environmental disturbance.

The rest of this work is organized as following. In Sec.~\ref{sec:three_level}, we will briefly review the STIRAP charging protocol for the three-level quantum battery. Then we introduce the STA technique via the counter-diabatic driving. In Sec.~\ref{sec:Gaussian}, the conventional STIRAP and the STA protocols are compared with each other in terms of the stored energy and the charging speed. In the numerical simulation, the transition lasers for both protocols are performed by the same type of Gaussian pulses and the counter-diabatic term can be derived analytically. The calculation in appendix~\ref{app:other_pulse} justifies that the result in the main text is irrelevant to the shape of laser pulse. In Sec.~\ref{sec:Rydberg}, we propose an implementation of our STA protocol for the three-level quantum battery in a Rydberg atomic system. In Sec.~\ref{sec:open_system}, we consider the decoherence effect to the charging process with practical parameters to demonstrate the robustness of our scheme against the environmental noise. In Sec.~\ref{sec:con}, we provide a discussion and then draw a conclusion.

\section{Quantum battery on three-level system}\label{sec:three_level}

A non-degenerate $n$-level quantum battery can be described by the Hamiltonian
\begin{equation}\label{eq:H0}
    H_0=\sum_{j=1}^n \varepsilon_j|\varepsilon_j\rangle\langle\varepsilon_j|,
\end{equation}
where $\varepsilon_j$'s are the eigenenergies of the bare system ordered by $\varepsilon_1<\varepsilon_2<\cdots<\varepsilon_n$. The internal energy of such a battery is given by ${\rm Tr}(\rho H_0)$, where $\rho$ is the density matrix. Charging such a quantum battery means that the internal energy increases when its state varies from $\rho$ to $\rho'$, i.e., ${\rm Tr}[(\rho'-\rho)H_0]\ge0$. The discharging process corresponds to the opposite variation. The charging dynamics is conventionally driven by the Hamiltonian $H=H_0+V(t)$, where the time-dependent part $V(t)$ describes the Hamiltonian of the external resonant or quasi-resonant driving field satisfying,
\begin{equation}\label{eq:Int}
    V(t)\neq 0, \quad {\rm for} \quad 0<t<\tau_c
\end{equation}
with charging time $\tau_c$. The stored energy at the moment $t$ is subsequently written as
\begin{equation}\label{eq:W}
    W(t)\equiv{\rm Tr}[\rho(t)H_0]-{\rm Tr}(\rho_0H_0).
\end{equation}
The initial state $\rho_0$ is conventionally defined as the passive state,
\begin{equation}\label{eq:passive}
\rho_0=\sum_{j=1}^n s_j |\varepsilon_j\rangle\langle\varepsilon_j| \quad {\rm with} \quad s_{j+1}\le s_j.
\end{equation}
Then the maximal stored energy named Ergotropy~\cite{bat1pioneer} can be attained when the quantum battery evolves from a passive state eventually to the highest level, $|\varepsilon_n\rangle$. The charging speed, also named charging power, can be averagely defined as
\begin{equation}\label{eq:P}
    P(t)=\frac{W(t)}{t},
\end{equation}
or can be instantaneously defined as $P=dW(t)/dt$~\cite{bat12fast,bat16dark,bat15fluc}. Here we adopt the average-value definition in Eq.~(\ref{eq:P}) as the charging speed~\cite{bat4chargemediate,bat5spinchain,bat8chargemediate,bat21cascade}, which is used to evaluate the performance of a quantum battery with respect to the time for full-charging.

In our three-level-system protocol for quantum battery, the ergotropy is simply the energy gap between the ground state $|\varepsilon_1\rangle$ and the second excited state $|\varepsilon_3\rangle$~\cite{bat13stable},
\begin{equation}\label{eq:Wmax}
    W_{\rm max}=\varepsilon_3-\varepsilon_1=\hbar(\omega_3-\omega_1).
\end{equation}
And the average charging speed can be represented by a dimensionless form,
\begin{equation}\label{eq:aP}
    P(\tau_c)\equiv\frac{W(\tau_c)/W_{\rm max}}{\Omega_0\tau_c},
\end{equation}
where $\tau_c$ is the full-charging period and $\Omega_0$ is the unit strength of the laser power. From now on, we set $\hbar=1$.

\subsection{Charging Hamiltonian for conventional STIRAP}

As demonstrated in Fig.~\ref{fig:demo}(c), the Hamiltonian of the three-level system interacting with the external driving fields can be written as~\cite{Semi_H1,Semi_H2,Semi_H3},
\begin{equation}\label{eq:Ht}
     H(t)=H_0+V(t),
\end{equation}
where the bare Hamiltonian of the system is
\begin{equation}\label{eq:H0_c}
    H_0=\omega_2|\varepsilon_2\rangle\langle\varepsilon_2|+\omega_3|\varepsilon_3\rangle\langle \varepsilon_3|
\end{equation}
with the ground-level energy $\omega_1\equiv0$ and the driving Hamiltonian is
\begin{equation}\label{eq:Hint}
V(t)=\sum_{j=1}^2\left[\Omega_j(t)e^{i\omega_{j,j+1}t}|\varepsilon_j\rangle\langle\varepsilon_{j+1}|\right]+{\rm H.c.}
\end{equation}
with the laser powers $\Omega_j$ and frequencies $\omega_{j,j+1}$, $j=1,2$. In the rotating frame with respect to
\begin{equation}\label{eq:Uni}
U_0(t)=\exp\left\{i\left[\omega_{12}|\varepsilon_2\rangle\langle\varepsilon_2|+\left(\omega_{12}+\omega_{23}\right)|\varepsilon_3\rangle\langle \varepsilon_3|\right]t\right\},
\end{equation}
the full Hamiltonian can be rewritten as,
\begin{eqnarray}\nonumber
H'(t)&=&U_0(t)H(t)U^\dagger_0(t)-iU_0(t)\dot{U}^\dagger_0(t) \\  \label{eq:H_dia}
&=&\begin{pmatrix}
    0 & \Omega_1(t) & 0 \\ \Omega_1(t) & \Delta & \Omega_2(t) \\ 0 & \Omega_2(t) & 0
     \end{pmatrix},
\end{eqnarray}
where $\Delta=\omega_2-\omega_{12}$. Since the non-vanishing detuning $\Delta$ does not give rise to distinct results in quality for both STIRAP and CD driving, we then stick to the one-photon-resonant condition, $\Delta=0$, in this work. Therefore, the frequencies of the the two laser pulses satisfy $\omega_2=\omega_{12}$ and $\omega_3-\omega_2=\omega_{23}$. The eigen-structure of the Hamiltonian in Eq.~(\ref{eq:H_dia}) thus reads,
\begin{equation}\label{eq:dark}
|\lambda_0=0\rangle=\begin{pmatrix}
\cos{\theta} \\ 0 \\ -\sin{\theta}
\end{pmatrix}, \quad |\lambda_{\pm}=\pm\Omega(t)\rangle=\frac{1}{\sqrt{2}}\begin{pmatrix}
\sin{\theta} \\ \pm 1 \\ \cos{\theta}
\end{pmatrix},
\end{equation}
where $\tan\theta=\Omega_1(t)/\Omega_2(t)$ and $\Omega(t)=\sqrt{\Omega_1^2(t)+\Omega_2^2(t)}$. The first eigenstate $|\lambda_0\rangle$ in Eq.~(\ref{eq:dark}) is the dark state that can be tuned in parametric space to adiabatically evolve from the initial state $|\varepsilon_1\rangle$ to the final state $|\varepsilon_3\rangle$.

In particular, the three-level system is initially prepared in the ground state and the boundary conditions for the driving pulses are set as $\Omega_1(0)\approx0$ and $\Omega_2(0)\neq 0$. With these functions of laser power adiabatically tuned such that $\Omega_1(\tau_c)\neq 0$ and $\Omega_2(\tau_c)\approx0$, the dark state will gradually approach $|\varepsilon_3\rangle$ without any population over $|\varepsilon_1\rangle$ and $|\varepsilon_2\rangle$. Then the quantum battery is fully charged in the end of this adiabatic charging process, which is stable without the precise control of the interaction time $\tau_c$, provided that $\Omega_1(t)\neq 0$ and $\Omega_2(t)\approx0$ with $t\geq\tau_c$. Even if one does not turn off the external interaction with charger, i.e., $V(t)\neq0$ when $t>\tau_c$, the quantum battery is still in the fully charged state $|\varepsilon_3\rangle$, which remains as the dark state for $t>\tau_c$. On the other hand, if the driving pulses are turned off, i.e., $V(t)=0$ when $t>\tau_c$, the Hamiltonian governing the dynamics of three-level quantum battery turns out to be the bare Hamiltonian $H_0$ in Eq.~(\ref{eq:H0_c}). In this case, since the fully charged state is the eigenstate of $H_0$, the quantum battery will also remain as $|\varepsilon_3\rangle$. Moreover, since $\lambda_0=0$ and $\langle\lambda_0|\dot{\lambda}_0\rangle=0$, the oscillating behavior of the gained energy $W(t)$ is avoided due to the fact that no quantal phase is accumulated during the evolution.

The preceding STIRAP charging protocol relies on a sufficient long charging time $\tau_c$ demanded by the adiabatic approximation. Otherwise for a short passage time $\tau_c$, transitions become inevitable between $|\lambda_0\rangle$ and $|\lambda_{\pm}\rangle$ and the middle level $|\varepsilon_2\rangle$ will be populated. As a result, the quantum battery cannot be fully charged leading thus to a low charging efficiency.

\begin{figure*}[htbp]
\centering
\subfigure{
\begin{minipage}{0.3 \textwidth}
\centering
\includegraphics[scale=0.28]{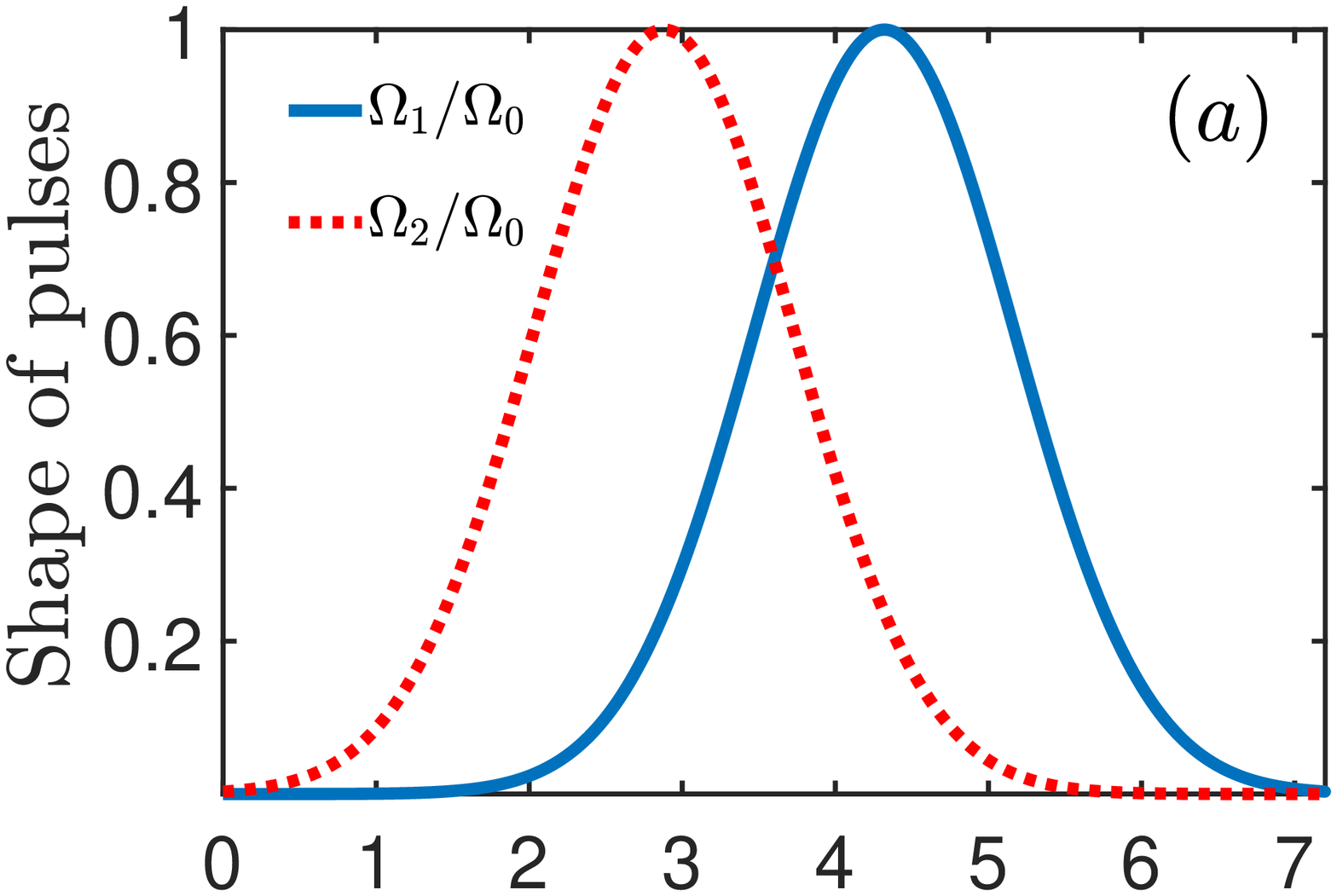}
\end{minipage}
}
\subfigure{
\begin{minipage}{0.3 \textwidth}
\centering
\includegraphics[scale=0.28]{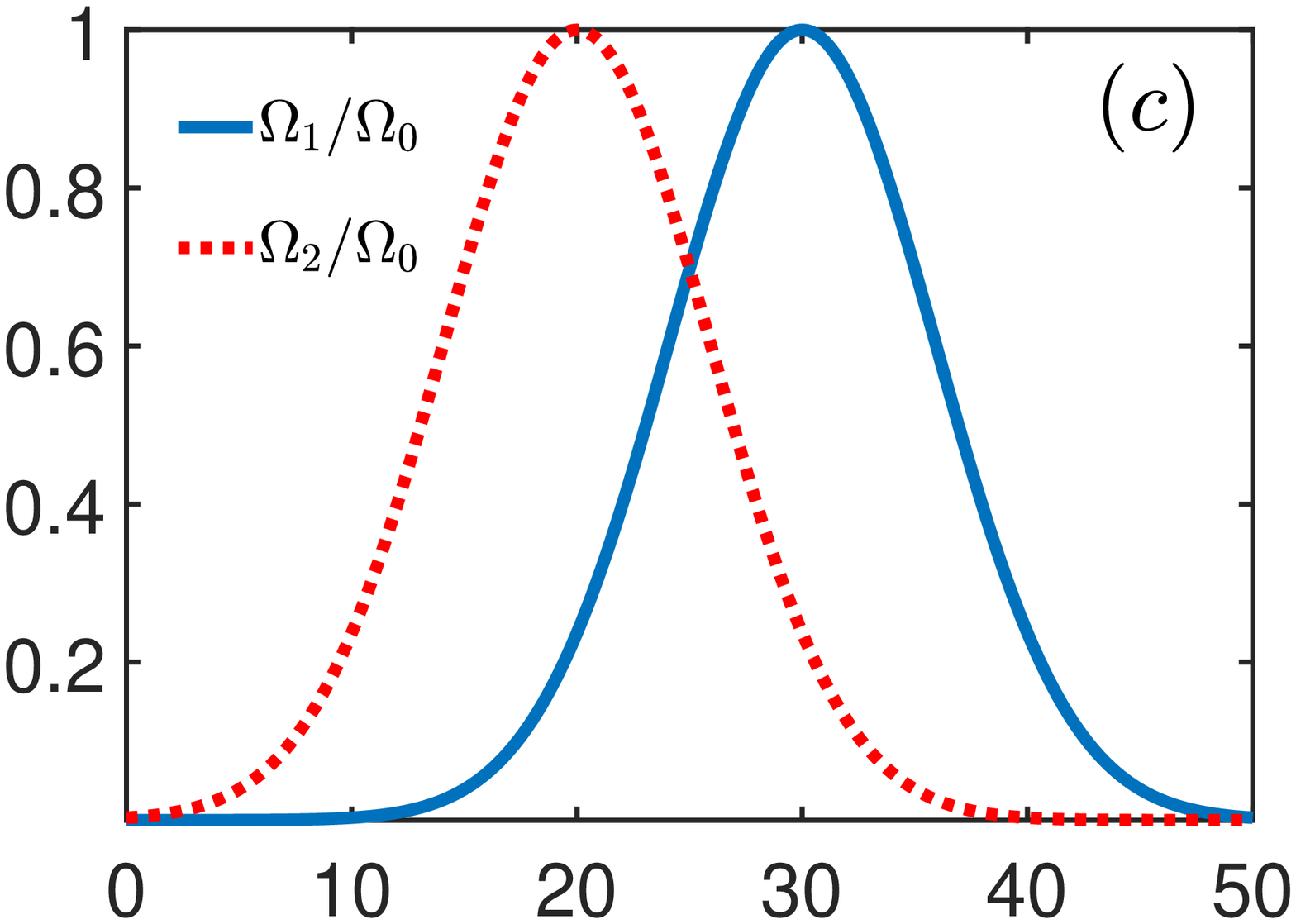}
\end{minipage}
}
\subfigure{
\begin{minipage}{0.3\textwidth}
\centering
\includegraphics[scale=0.28]{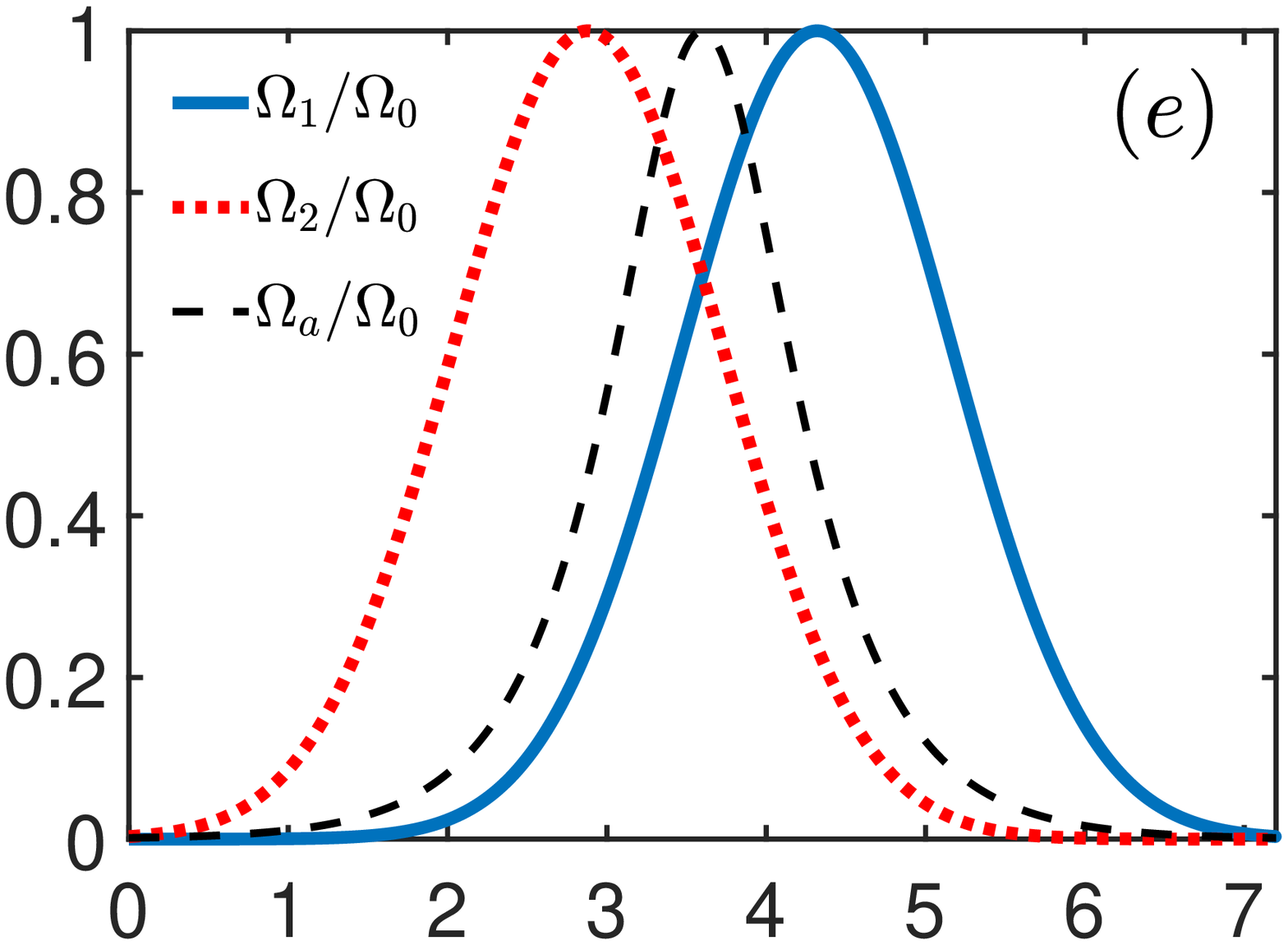}
\end{minipage}
}
\subfigure{
\begin{minipage}{0.3\textwidth}
\centering
\includegraphics[scale=0.28]{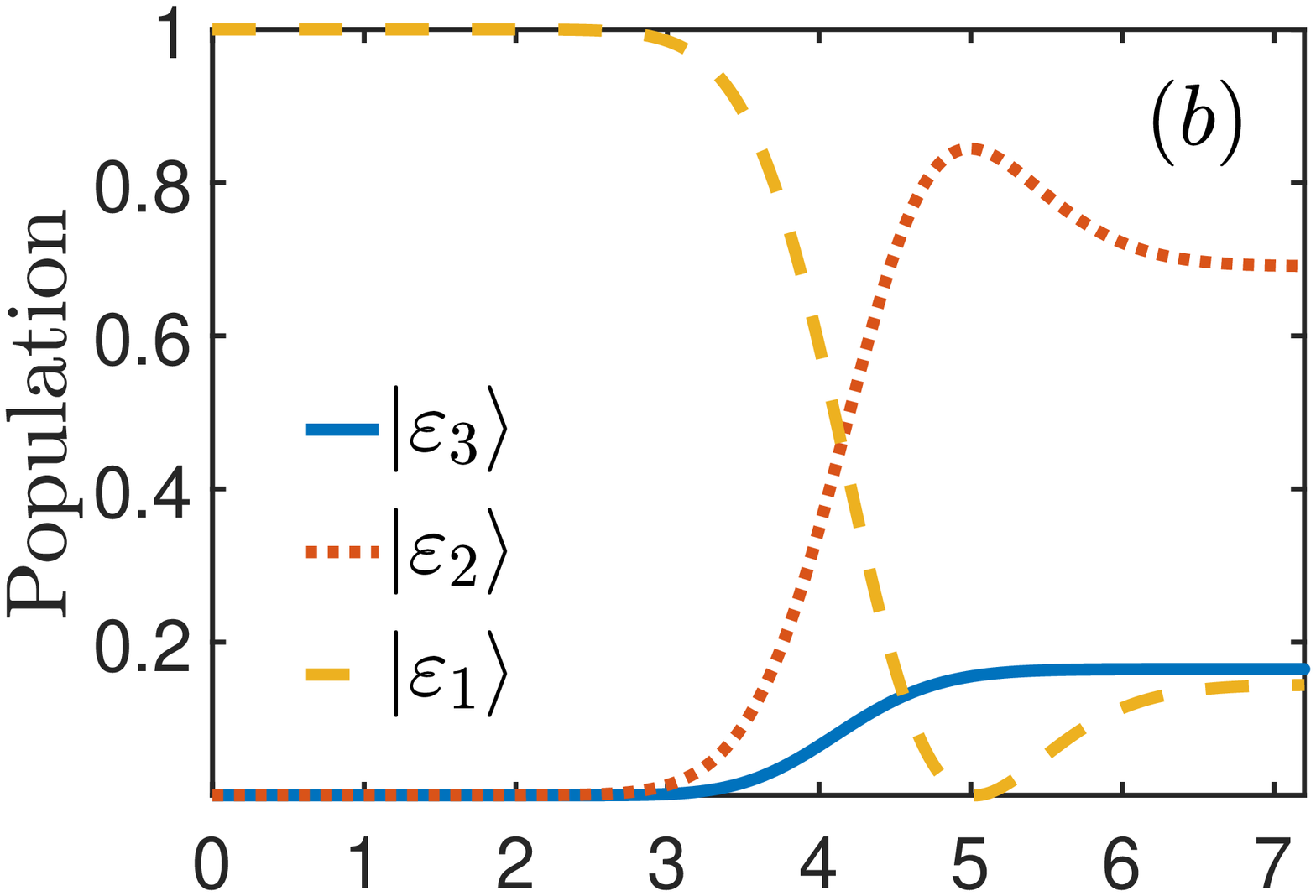}
\end{minipage}
}
\subfigure{
\begin{minipage}{0.3\textwidth}
\centering
\includegraphics[scale=0.28]{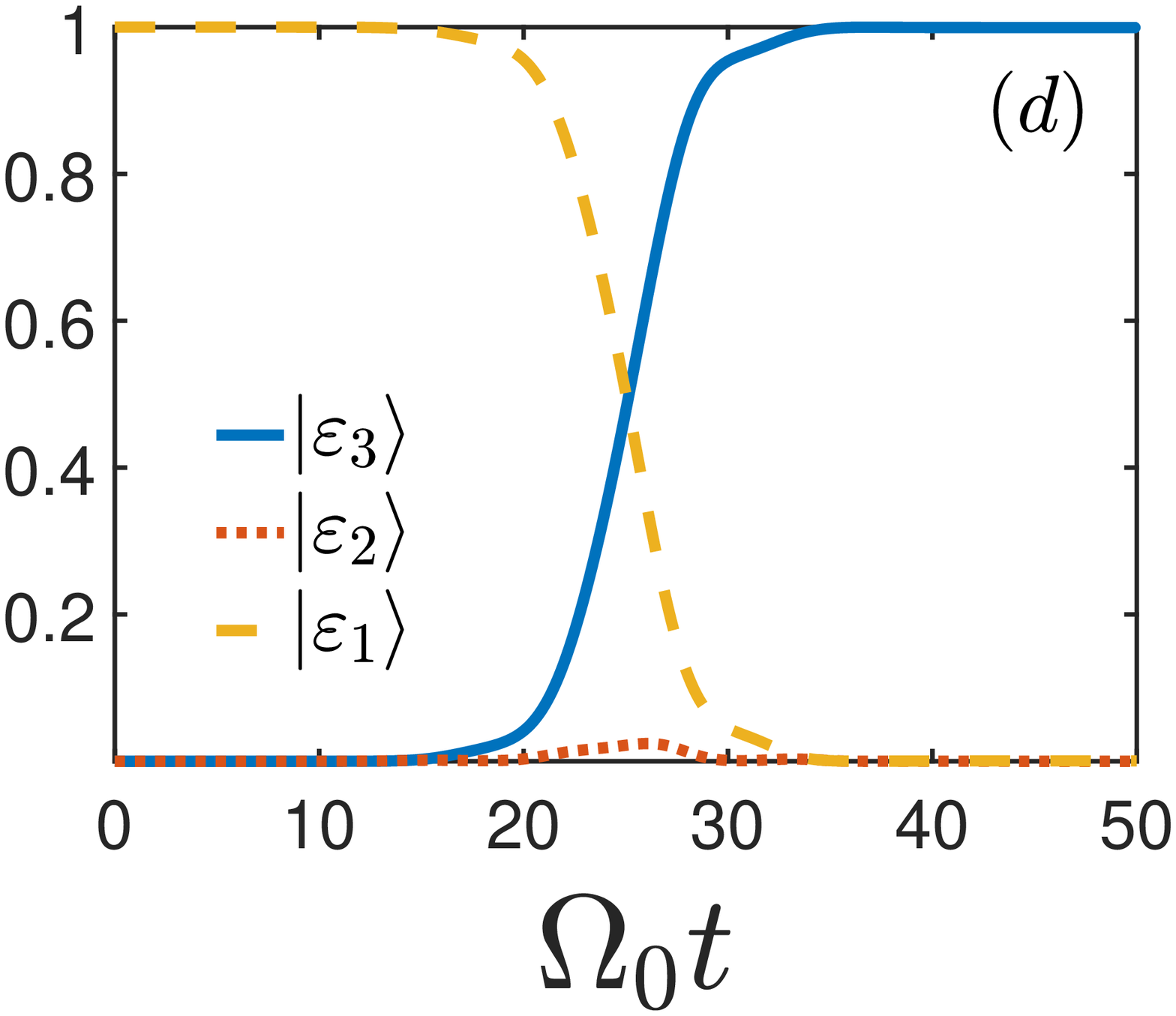}
\end{minipage}
}
\subfigure{
\begin{minipage}{0.3\textwidth}
\centering
\includegraphics[scale=0.28]{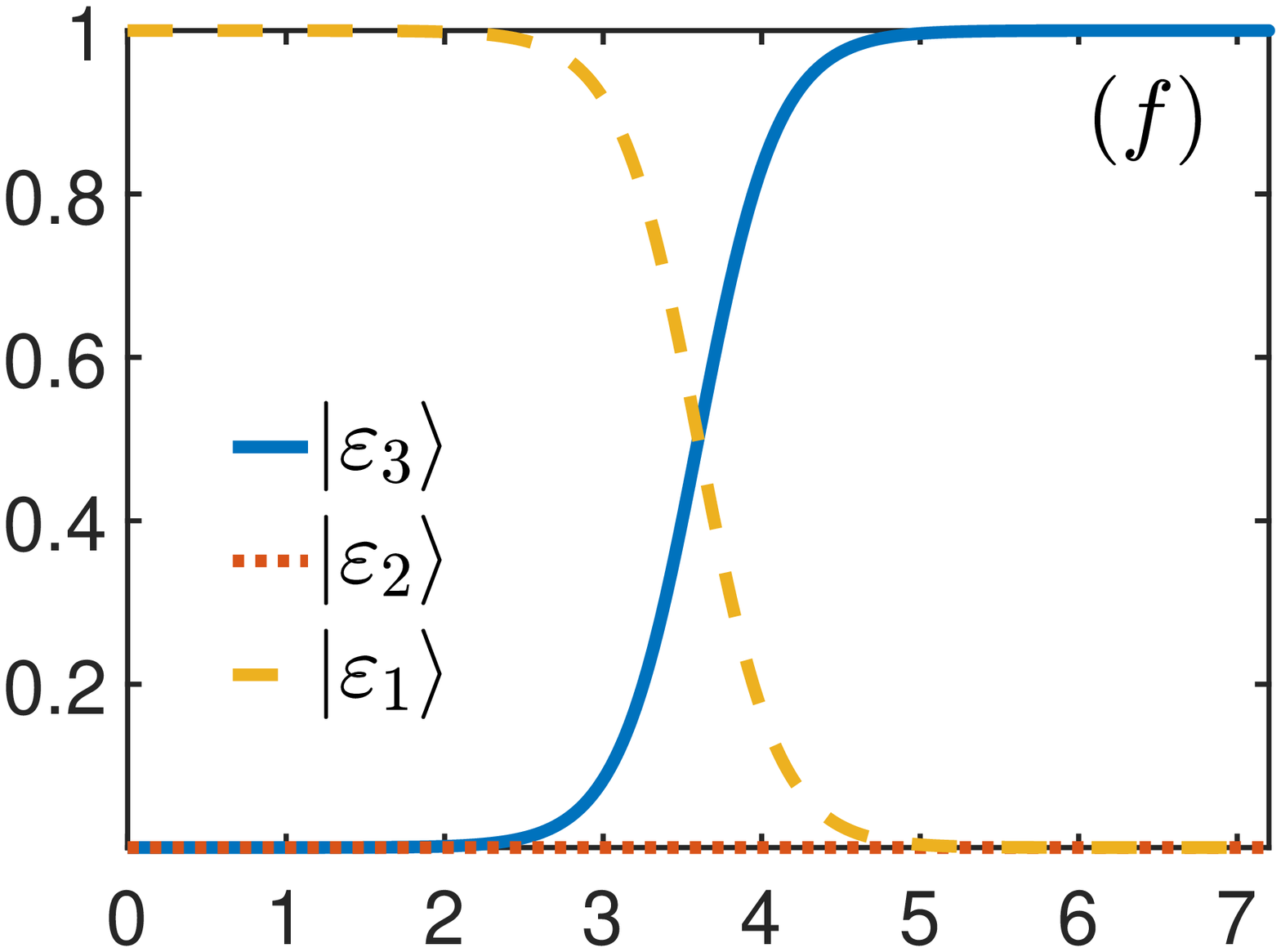}
\end{minipage}
}
\caption{(Color online) Pulse shapes for the charging protocols of STIRAP (a), (c) and STA (e) and the induced population dynamics for the three-level quantum battery (b), (d) and (f) by the corresponding pulses. The charging period for (a), (b) and (e), (f) is chosen as $\Omega_0\tau_c=7.2$, and that for (c), (d) is chosen as $\Omega_0\tau_c=50$. In (a), (c) and (e), $\Omega_1(t)$ and $\Omega_2(t)$ are plotted with the solid-blue lines and the dotted-red lines, respectively. In (e), $\Omega_a(t)$ is plotted with the dashed-black line. In (b), (d) and (f), the populations over $|\varepsilon_1\rangle$, $|\varepsilon_2\rangle$, and $|\varepsilon_3\rangle$ are plotted with the yellow-dashed lines, the dotted-red lines and the blue-solid lines, respectively. The charging process is completed when the population over $|\varepsilon_3\rangle$ becomes stable with time. } \label{fig:dynamics}
\end{figure*}

\subsection{Charging Hamiltonian for CD driving}

To suppress the transitions between various instantaneous eigenstates for an adiabatic charging process and to accelerate the charging speed, one can apply the counter-diabatic term to facilitate the system to follow the preceding adiabatic path, i.e., $|\lambda_0\rangle$ in Eq.~(\ref{eq:dark}), with a potentially much shorter $\tau_c$.

According to the standard recipe in Eq.~(\ref{eq:HCD}), $H_{\rm CD}$ in this work can be explicitly expressed in the basis spanned by $|\varepsilon_1\rangle$, $|\varepsilon_2\rangle$ and $|\varepsilon_3\rangle$ as,
\begin{subequations}\label{eq:HCD_c}
\begin{equation}
H_{\rm CD}(t)=i\begin{pmatrix}
    0 & 0 & \Omega_a(t) \\ 0 & 0 & 0 \\ -\Omega_a(t) & 0 & 0
    \end{pmatrix},
\end{equation}
where
\begin{equation}\label{eq:Oa}
    \Omega_a(t)=\dot{\theta}=\frac{\dot{\Omega}_1(t)\Omega_2(t)-\Omega_1(t)\dot{\Omega}_2(t)}{\Omega^2(t)}.
\end{equation}
\end{subequations}
Combining the assistant Hamiltonian in Eq.~(\ref{eq:HCD_c}) with the conventional STIRAP Hamiltonian in Eq.~(\ref{eq:H_dia}), one can obtain a Hamiltonian for shortcut to adiabaticity,
\begin{eqnarray}\nonumber
     H_{\rm STA}(t)&=&H'(t)+H_{\rm CD}(t) \\ \label{eq:HSTA}
     &=&\begin{pmatrix}
    0 & \Omega_1(t) & i\Omega_a(t) \\ \Omega_1(t) & \Delta & \Omega_2(t) \\ -i\Omega_a(t) & \Omega_2(t) & 0
    \end{pmatrix}.
\end{eqnarray}
As will be shown in the following section, the evolution time $\tau_c$ required for the full-charging process can be much shorter than the conventional STIRAP protocol.

\section{Charging process of STA protocol}\label{sec:Gaussian}

In this section, we present the charging dynamics under either conventional STIRAP or our STA protocols in terms of stored energy and charging speed. It is found that the shape of the laser pulse is irrelevant to the charging process (see appendix~\ref{app:other_pulse}). Here we use the Gaussian pulses for the two driving pulses $\Omega_{1,2}(t)$, which are popularly performed in the existing works for STIRAP~\cite{kumar2016stimulated,Chao2016STA}. And the counter-diabatic pulse $\Omega_a(t)$ can be analytically derived. Specifically, we have
\begin{equation}\label{eq:O1t}
    \Omega_{1}(t)=\Omega_0\exp\left[-\frac{\left(t-\tau_c/2-\alpha\right)^2}{\sigma^2}\right],
\end{equation}
\begin{equation}\label{eq:O2t}
    \Omega_{2}(t)=\Omega_0\exp\left[-\frac{\left(t-\tau_c/2+\alpha\right)^2}{\sigma^2}\right],
\end{equation}
where $\Omega_0$ is the unit strength of pulses. In the following numerical simulations, the shape parameters are set as $\alpha=\tau_c/10$ and $\sigma=\tau_c/6$ to approximately meet the boundary conditions for the adiabatic passage of the dark state. Given the expression of $\Omega_1(t)$ and $\Omega_2(t)$, one can explicitly obtain the assistant pulse for the CD term
\begin{equation}\label{eq:Oat}
    \Omega_a(t)=\frac{2\alpha}{\sigma^2}\cosh^{-1}\left(\frac{4\alpha t-2\alpha\tau_c}{\sigma^2}\right),
\end{equation}
according to Eq.~(\ref{eq:Oa}). Note now $\tau_c$ becomes a free variable. In order to compare the charging performance between these two protocols, one can restrict the peak value of $\Omega_a(t)$ to be equivalent to $\Omega_0$, the unit strength of the preceding pulses, by tuning $\tau_c$.

We first demonstrate the charging dynamics of STIRAP with driving pulses in Eqs.~(\ref{eq:O1t}) and (\ref{eq:O2t}) and STA with the extra CD driving pulse in Eq.~(\ref{eq:Oat}) by the evolution of populations over $|\varepsilon_j\rangle$, $j=1,2,3$, against the dimensionless time $\Omega_0t$. The initial state of the quantum battery is assumed to be the empty state $|\psi(0)\rangle=|\varepsilon_1\rangle$. The power of both STIRAP pulses and CD pulse are supposed to be comparable with each other for the sake of moderate cost in experiments. Surely the full-charging period can be further reduced by an overhead of a stronger power of $\Omega_a(t)$. If the power of the CD pulse is under the constraint of ${\rm Max}[|\Omega_a(t)|]\leq\Omega_0$, then the lower-bound of charging period of STA protocol is found to be $\Omega_0\tau_c\approx7.2$. With this charging time, the STIRAP protocol definitely fails since the asymptotic population over $|\varepsilon_3\rangle$ is less than $0.2$, as shown in Fig.~\ref{fig:dynamics}(b). Clearly this process is not adiabatic because the two driving pulses $\Omega_1(t)$ and $\Omega_2(t)$ is not sufficiently slowly-varying with their parameters in such a short time interval. The population is thus not fully transferred from the ground state $|\varepsilon_1\rangle$ to the fully charged state $|\varepsilon_3\rangle$ and the middle state $|\varepsilon_2\rangle$ is significantly populated.

A practical adiabatic passage can be realized by sufficiently extending the shape of pulses. We show in Fig.~\ref{fig:dynamics}(c) and (d) that by setting the charging time $\Omega_0\tau_c=50$, the middle state $|\varepsilon_2\rangle$ is little populated around the moment $\Omega_0t\approx26$ and the state $|\varepsilon_3\rangle$ is fully populated after the state of the quantum battery becomes stable. However, this process requires a long evolution time and might be destroyed in the presence of quantum decoherence. To hold the same charging performance within a much shorter charging time, one can employ the CD pulse to realize the STA protocol. In Fig.~\ref{fig:dynamics}(e) and (f), the population over the empty state transfers completely to the fully charged state $|\varepsilon_3\rangle$ through the adiabatic passage of the dark state in Eq.~(\ref{eq:dark}). Note that during the charging process, the middle state is never populated. Comparing Fig.~\ref{fig:dynamics}(d) and (f), the STA protocol is faster than the conventional STIRAP by almost one order in magnitude.

\begin{figure}[htbp]
\centering
\includegraphics[width=0.4\textwidth]{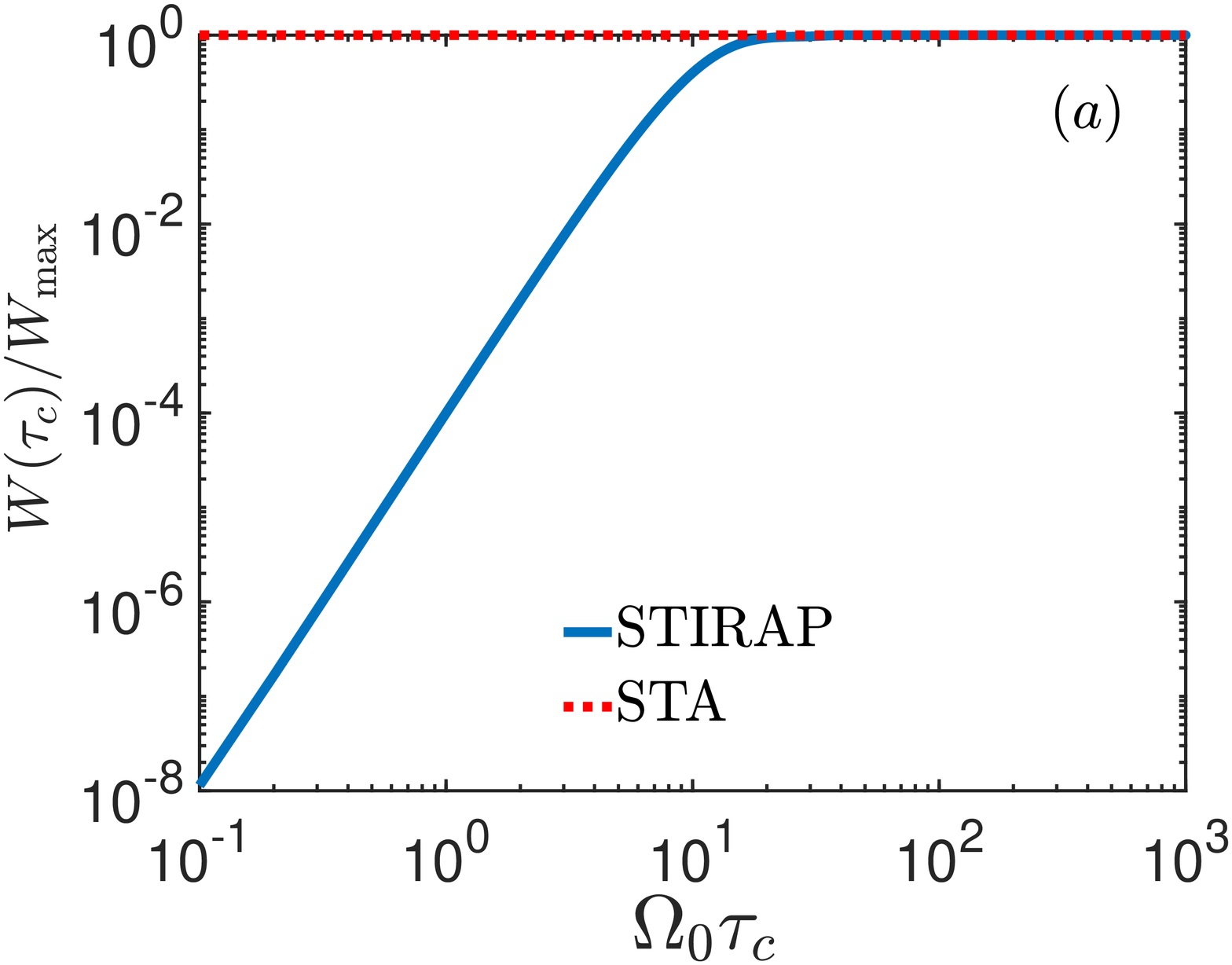}
\includegraphics[width=0.4\textwidth]{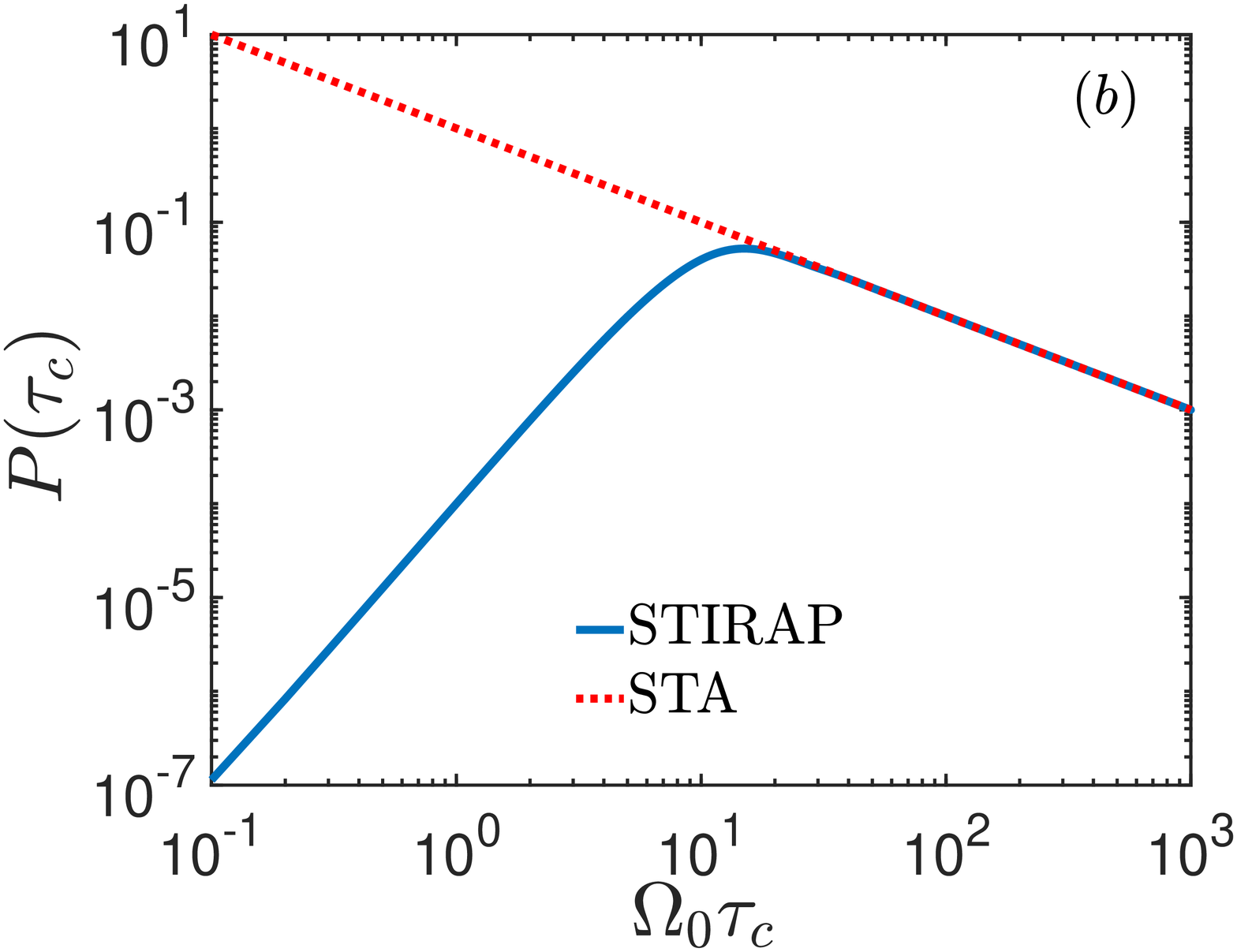}
\caption{(Color online) (a) Normalized stored energy and (b) average charge speed as functions of charging time $\tau_c$ under the STIRAP (blue-solid lines) and STA (red-dotted lines) charging protocols. } \label{fig:WP}
\end{figure}

Next we evaluate the stored energy $W(\tau_c)$ in Eq.~(\ref{eq:W}) and the charging speed $P(\tau_c)$ in Eq.~(\ref{eq:aP}) to explicitly demonstrate the advantage of our STA charging protocol over the conventional STIRAP protocol, given the pulses in Eqs.~(\ref{eq:O1t}), (\ref{eq:O2t}) and (\ref{eq:Oat}). One can observe that in Fig.~\ref{fig:WP}(a), when the charging time is as short as $\Omega_0\tau_c=0.1$, the STA charging protocol still works for the quantum battery. And by contrast, the normalized stored energy under the STIRAP protocol is approximately lower by $8$ orders in magnitude. The stored energy $W(t)$ by STIRAP is rapidly enhanced with the increasing charging time $\tau_c$. Eventually it attains $W_{\rm max}$ and remains stable when $\Omega_0\tau_c$ is over $20$, which could be regarded the adiabatic limit, i.e., the lower-bound of the evolution time for STIRAP under the fixed parameters in Eqs.~(\ref{eq:O1t}) and (\ref{eq:O2t}). Thus the STA protocol is more preferable than the STIRAP protocol when the charging time is shorter than this limit.

Figure.~\ref{fig:WP}(b) demonstrates the distinction of the average charge speeds between these two charging processes. Here we employ the dimensionless charge speed defined in Eq.~(\ref{eq:aP}). As expected, the charging speed of STA shows a monotonic-decreasing behavior since the overall charging time is enlarging while the stored energy remains as $W_{\rm max}$. In contrast, the charging speed of STIRAP is determined by the increasing stored energy and charging time. The charging speeds of the two protocols are distinct by several orders in magnitude. It is observed that when $\Omega_0\tau_c\approx1$, the charging speed of STA is almost $10^3$ times as that of conventional STIRAP. The latter gradually increases with $\tau_c$ until it starts to be coalescent with the speed line for STA at the point of adiabatic limit. When the requirement of strong counter-diabatic driving pulse is relieved to be not larger than $\Omega_0$, the charging speed of STA is still about multiple times as that of STIRAP. For example, when $\Omega_0\tau_c=7.2$, the lower-bound of charging time obtained by $\Omega_a(t)\leq\Omega_0$, the dimensionless average charge speed assisted by the CD pulse is about $0.139$ and in contrast, the speed of conventional STIRAP is merely $0.023$. This result is consistent with the result of stored energy as in Fig.~\ref{fig:WP}(a).

It is worth emphasizing that the preceding results about the properties of the quantum battery under STA protocol is independent on the choice of pulse shapes. This argument is supported by the appendix~\ref{app:other_pulse}, in which similar accelerating effects by virtue of the counter-diabatic driving can be found with sinusoid pulse and ramp-like pulse. We therefore are able to summarize that in principle, the STA technique inherits all the advantage of the conventional STIRAP protocol, such as the adiabatic passage of the dark state and the stable charging in three-level system. Moreover, the charging process can be speeded up with the overhead of the extra CD driving pulse.

\section{Physical Implementation}\label{sec:Rydberg}

The preceding charging protocol based on a cascade-type three-level system can be implemented in various experimental systems, including the superconducting circuit~\cite{Cascade1tansmon,Cascade2tansmon,Cascade3supercon,Cascade4supercon}, trapped ion~\cite{Cascade5ion}, and Rydberg atoms~\cite{Cascade6Rydberg,Cascade7Rydberg,Cascade8Rydberg}, to name a few. In this section, we establish our STA quantum battery on the Rydberg atoms of $^{79}$Rb under proper optical driving. It is an accessible platform to implement a three-level system with cascade configuration. According to Refs.~\cite{Cascade7Rydberg} and~\cite{Saffman2010review}, the lifetime of Rydberg atoms approaches $42.3\mu$s when $n=43$ ($n$ is the principle quantum number) and scales as $n^3$ at the room temperature.

\begin{figure}[htbp]
    \centering
    \includegraphics[width=0.45\textwidth]{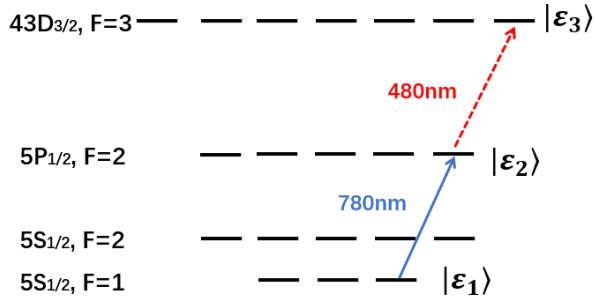}
    \caption{(Color online) Atomic level structure and driving lasers for the charging on the Rydberg atoms of $^{79}$Rb. The levels $5S_{1/2}(F=1)$, $5P_{1/2}(F=2)$ and $43D_{3/2}(F=3)$ are chosen as the our quantum-battery levels $|\varepsilon_1\rangle$, $|\varepsilon_2\rangle$, and $|\varepsilon_3\rangle$, respectively. The $780$nm laser (blue-solid line) and the $480$nm laser (red-dotted line) are used to play the roles of driving pulses $\Omega_1(t)$ and $\Omega_2(t)$ for STIRAP, respectively. }  \label{fig:Rb}
\end{figure}

In particular, we choose the coherent transitions connecting the levels $5S_{1/2}(F=1)-5P_{1/2}(F=2)-43D_{3/2}(F=3)$ in $^{79}$Rb as our three-level quantum battery [see Fig.~\ref{fig:Rb}], where $5S_{1/2}(F=1)$, $5P_{1/2}(F=2)$ and $43D_{3/2}(F=3)$ are $|\varepsilon_1\rangle$, $|\varepsilon_2\rangle$, and $|\varepsilon_3\rangle$, respectively. According to Ref.~\cite{Cascade7Rydberg}, the laser pulses interacting with the Rydberg atom, i.e., the $780$nm laser in charge of the $5S_{1/2}(F=1)-5P_{1/2}(F=2)$ transition and the $480$nm laser in charge of the $5P_{1/2}(F=2)-43D_{3/2}(F=3)$ transition, can be used as the two driving pulses in our charging protocol. They satisfy the one-photon resonant condition, i.e., $\Delta=0$.

To construct a STA protocol, the assistant counter-diabatic Hamiltonian is demanded to accelerate the charging speed. Theoretically, CD term in Eq.~(\ref{eq:HCD_c}) describes the transition between the ground state $|\varepsilon_1\rangle$ and the second excited state $|\varepsilon_3\rangle$, which however is forbidden for a natural cascade-three-level atom. Rather than directly applying a third pulse to the Rydberg atom, we can effectively simulate and supplement the CD term by two modified stimulated Ramen pulses, which can be obtained by a similar transformation over the full STA Hamiltonian in Eq.~(\ref{eq:HSTA}). Note it can be decomposed by the generators for the SU$(2)$ group~\cite{Chao2016STA}:
\begin{equation}\label{eq:HSTA_c}
    H_{\rm STA}(t)=\Omega_1(t)S_1+\Omega_2(t)S_3-\Omega_a(t)S_2,
\end{equation}
where
\begin{equation}
S_1=\begin{pmatrix}
0 & 1&0\\1 & 0 &0\\0&0&0
\end{pmatrix}, \quad S_2=\begin{pmatrix}
0 & 0&-i\\0 & 0 &0\\i&0&0
\end{pmatrix}, \quad S_3=\begin{pmatrix}
0 & 0&0\\0 & 0 &1\\0&1&0
\end{pmatrix}.
\end{equation}
They satisfy the commutation relation $[S_i,S_j]=-i\varepsilon_{ijk}S_k$ with $i,j,k=1,2,3$. With respect to the transformation,
\begin{equation} \label{eq:Uni2}
\begin{aligned}
     U(t)&=e^{-i\phi(t)S_3}=\begin{pmatrix}
    1 & 0 & 0 \\ 0 & \cos{\phi(t)} & -i\sin{\phi(t)}\\ 0 & -i\sin{\phi(t)} & \cos{\phi(t)}
    \end{pmatrix},
\end{aligned}
\end{equation}
where $\phi(t)$ is an undetermined time-dependent parameter, one can have,
\begin{equation}\label{eq:HSTA_r}
    \begin{aligned}
        H'_{\rm STA}&=U^\dagger H U -i U^\dagger \dot{U} \\
        &=\tilde{\Omega}_1(t)S_1+\tilde{\Omega}_2(t)S_3-\tilde{\Omega}_aS_2,
    \end{aligned}
\end{equation}
where the modified pulse strengthes have the form as following,
\begin{eqnarray}\nonumber
    \tilde{\Omega}_1(t)&=&\Omega_1(t)\cos{\phi}(t)+\Omega_a(t)\sin{\phi}(t), \\ \nonumber
    \tilde{\Omega}_2(t)&=&\Omega_2(t)-\dot{\phi}(t), \\
    \tilde{\Omega}_a(t)&=&\Omega_1(t)\sin{\phi}(t)-\Omega_a(t)\cos{\phi}(t).
\end{eqnarray}
To eliminate transition-forbidden term about $S_2$ in the cascade three-level quantum battery, we can set $\tilde{\Omega}_a(t)=0$. Then $\phi(t)$ is determined by
\begin{equation}\label{eq:phi}
    \tan{\phi(t)}=\frac{\Omega_a(t)}{\Omega_1(t)},
\end{equation}
leaving the rotating Hamiltonian consisted of two allowed transition lasers with the modified shape functions.
\begin{equation}\label{eq:HSTA_mr}
H'_{\rm STA}(t)=\tilde{\Omega}_1(t)S_1+\tilde{\Omega}_2(t)S_3.
\end{equation}
This effect of $H'_{\rm STA}$ in Eq.~(\ref{eq:HSTA_mr}) is equivalent to the original STA Hamiltonian in Eq.~(\ref{eq:HSTA}). We thus can effectively simulate the CD term in a cascade three-level system. The overhead is that the shape function might not be as regular as the original one.

\section{Decoherence impact on charging}\label{sec:open_system}

Now we consider the scenario in which the adiabatic charging process is exposed to quantum decoherence with the practical decay rates in the Rydberg atom as a possible platform for our STA protocol. In particular, we are interested in the charging dynamics governed by the phenomenological master equation~\cite{cohen1998atom,breuer2002theory},
\begin{equation}\label{eq:Lindblad}
\begin{aligned}
       \frac{\partial\rho}{\partial t}=&-i[H(t), \rho]\\
       &+\sum_{k=2,3}\left\{\frac{\gamma_{z,k}}{2}\mathcal{L}[\sigma^z_{kk}](\rho)
       +\frac{\gamma_{-,k}}{2}\mathcal{L}[\sigma^-_{k-1,k}](\rho)\right\},
\end{aligned}
\end{equation}
where $H(t)$ could be $H'(t)$ in Eq.~(\ref{eq:H_dia}) or $H_{\rm STA}(t)$ in Eq.~(\ref{eq:HSTA}) in order to compare the charging dynamics under different protocols, and the superoperation $\mathcal{L}[A](\rho)$ with respect to the system operator $A$ for decoherence channel is defined by
\begin{equation}\label{eq:Super_Op}
    \mathcal{L}[A](\rho)\equiv 2A\rho A^\dagger - A^\dagger A\rho-\rho A^\dagger A.
\end{equation}
In this cascade three-level system, the energy dissipation effects are described by $A=\sigma^-_{k-1,k}\equiv|\varepsilon_{k-1}\rangle\langle\varepsilon_k|$, $k=2,3$, implying the quantum jump from $|\varepsilon_k\rangle$ to $|\varepsilon_{k-1}\rangle$. And the phase decoherence can be described by $A=\sigma_{kk}^z\equiv|\varepsilon_k\rangle\langle\varepsilon_k|-|\varepsilon_{k-1}\rangle\langle\varepsilon_{k-1}|$ implying the fluctuations of the two excited states.

\begin{figure}[htbp]
\centering
\includegraphics[width=0.4\textwidth]{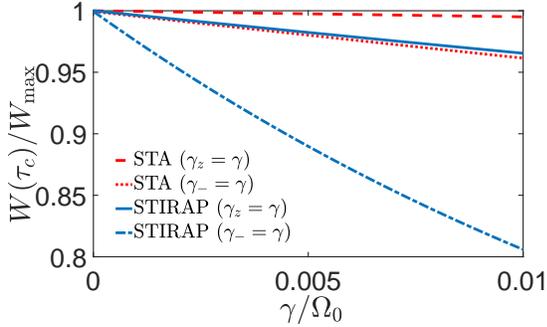}
\caption{(Color online) The normalized stored energy in the open quantum battery at a predesignated full-charging time $\Omega_0\tau_c$, which is set as $7.2$ and $50$ for the STA and the STIRAP protocols, respectively. The red-dotted line, the red-dashed line, the blue-dashed-dotted line, and the blue-solid represent the STA charging against dissipation, the STA charging against dephasing, the STIRAP protocol against dissipation, and the STIRAP protocol against dephasing, respectively. }\label{fig:open_system}
\end{figure}

To show the robustness of STA charging protocol against different decoherence effects, we focus individually on the dissipation and the dephasing processes in numerical simulation rather than considering their collective effect, i.e., when the dissipation (dephasing) channel is switched on, the dephasing (dissipation) rate is set as zero. For simplicity, both dephasing and dissipation rates are set to be independent of the involved states, i.e., $\gamma_{z,1}=\gamma_{z,2}=\gamma_z$ and $\gamma_{-,1}=\gamma_{-,2}=\gamma_-$. And to compare the robustness of both full-charging protocols, we set the total dimensionless charging time $\Omega_0\tau_c$ to be $7.2$ and $50$ for STA and STIRAP, respectively.

In Fig.~\ref{fig:open_system}, we plot the normalized stored energy $W(\tau_c)/W_{\rm max}$ as a function of $\gamma_z$ or $\gamma_-$ in four scenarios. All results for the final stored energy of the battery are expected to gradually decrease as the decay rates increase. One can observe that (1) for both decoherence channels, the STA protocol assisted by the CD driving (the red lines) is more robust than the conventional STIRAP protocol (the blue lines); (2) for both charging protocols, the destructive effect from the dissipation channel is significantly stronger than that from the dephasing channel. For example, when $\gamma_{-}/\Omega_0=0.01$, the stored energy of STIRAP under dissipation will decline to $0.806W_{\rm max}$ and that of STA is still about $0.961W_{\rm max}$. In the presence of pure dephasing, the stored energy of both STIRAP and STA at $\gamma_z/\Omega_0=0.01$ can be maintained over $0.965W_{\rm max}$. The former result relies heavily on the fact that the full-charging time of STA is much shorter than that of STIRAP. The latter can be understood that the pure dephasing effect does not directly influence the energy-charging process but slightly affect the adiabatic passage.

\begin{figure}[htbp]
\centering
\includegraphics[width=0.4\textwidth]{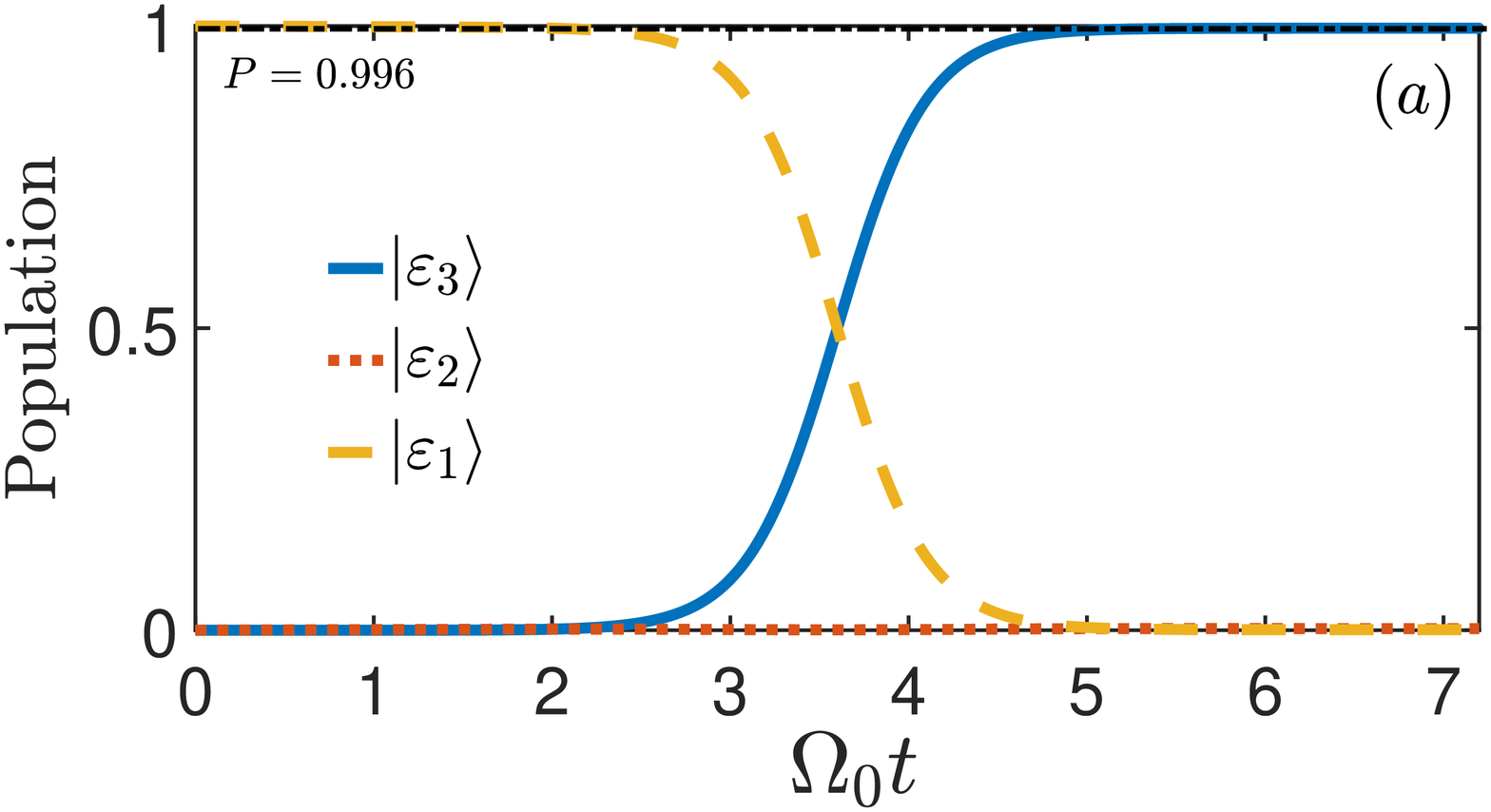}
\includegraphics[width=0.4\textwidth]{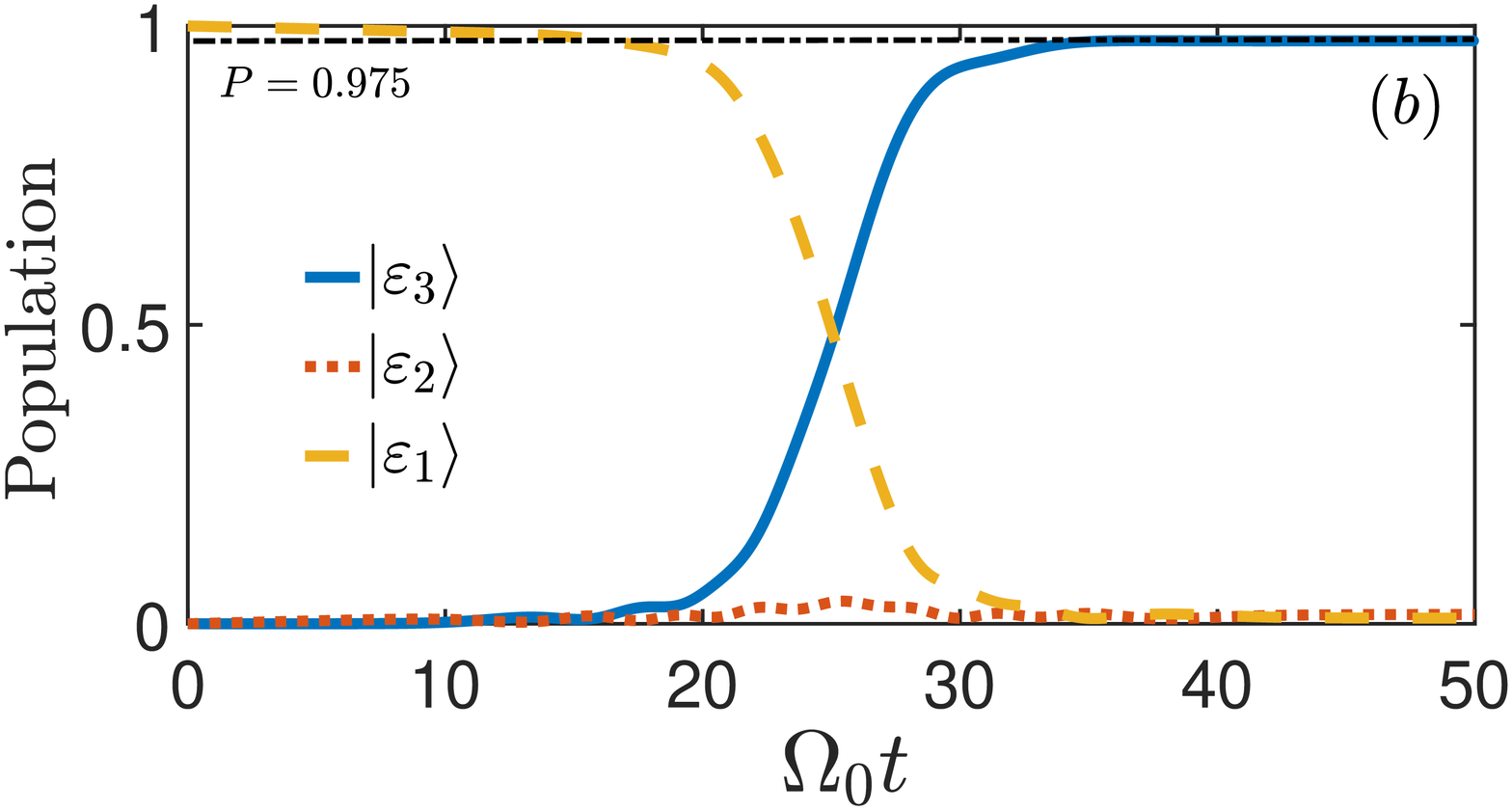}
\caption{(Color online) Population dynamics of the open quantum battery under (a) STA and (b) STIRAP protocols with a fixed dissipation rate $\gamma_-/\Omega_0=0.001$. Populations over $|\varepsilon_1\rangle$, $|\varepsilon_2\rangle$ and $|\varepsilon_3\rangle$ are plotted with the yellow-dashed line, the red-dotted line and the blue-solid line, respectively. The black dotted-dashed lines in both sub-figures evaluates the final population over $|\varepsilon_3\rangle$. }\label{fig:open_dynamics}
\end{figure}

Figure~\ref{fig:open_dynamics} is contributed to directly show the robustness of the STA protocol against quantum decoherence, where the dissipation rate is set as $\gamma_-/\Omega_0=0.001$. This choice is consistent with the condition in practical experiments in order of magnitude. For example, the Rabi frequency of the Rydberg atomic system ($^{79}$Rb) is about $1\sim100$ MHz~\cite{Cascade6Rydberg} and the spontaneous emission rate is about $10$ kHz.

Figure~\ref{fig:open_dynamics}(a) and (b) show the population dynamics of the battery under STA and STIRAP protocols, respectively. And their evolution times are chosen respectively the same as those of Fig.~\ref{fig:dynamics}(d) and (f). In comparison to the ideal charging process in Fig.~\ref{fig:dynamics}, the population over the fully-charged state $|\varepsilon_3\rangle$ under STA protocol is still nearly unit (about $0.996$) and nearly no population over the ground state $|\varepsilon_1\rangle$ as well as the middle state $|\varepsilon_2\rangle$ can be observed. For the STIRAP protocol, a small fraction of population will be transferred to the middle state $|\varepsilon_2\rangle$ in the end of charging, by which the stable population over $|\varepsilon_3\rangle$ is about $0.975$, a slightly lower than it for the STA protocol. Therefore, our STA protocol is also more favorable than the conventional STIRAP protocol in terms of charging dynamics.

\section{Discussion and conclusion}\label{sec:con}

The microscopic model for our quantum battery protocol on a shortcut to adiabaticity is chosen as a cascade three-level system as demonstrated in Sec.~\ref{sec:three_level}. It should be emphasized that this protocol can be willingly extended to other types of three-level systems. We can also construct such a STA battery with the $\Delta$-type (loop-type) three-level systems that are popularly seen in the superconducting qubit or qutrit systems, such as flux qubit~\cite{Orlando1999flux}, phase qubit~\cite{li2012dynamical,Martinis2002phase}, fluxonium qubit~\cite{manucharyan2009fluxonium,pop2014coherent}, and transmon qubit circuits~\cite{vepsalainen2019superadiabatic}. In the $\Delta$-type three-level system with no forbidden transition, the counter-diabatic driving can be directly performed in addition to the two stimulated Raman pulses. Consequently, one does not need to seek a special rotating frame to modify the existing driving pulses. It might be more convenient in experiments to attain a STA charging process. Thus our STA protocol is a significantly versatile method which can be applied in various systems. In Sec.~\ref{sec:Rydberg}, we discuss the application of the STA quantum battery in the Rydberg atomic system ($^{79}$Rb). We note here that Rydberg atoms are scalable in 2D optical lattice~\cite{Saffaman2008Scalability} and now have a prospect to be a building block for quantum network. One can then expect that the Rydberg-atom-based quantum battery can be further scaled up by virtue of manipulation technique on Rydberg blockage.

In this work, we propose a fast and stable charging protocol in a three-level quantum battery based on the adiabatic passage of the system dark-state. It is equivalent to a shortcut to adiabaticity assisted by an extra counter-diabatic driving. We have shown that a more efficient quantum battery can be established by the STA protocol in the light of the charging performance under various shapes of pulses. With a moderate overhead of the extra CD pulse, STA battery is able to accelerate the charging process by nearly one order in magnitude. Also, the robustness of the STA battery is measured by the impact from the effects of quantum dissipation and quantum dephasing on the charging protocols. Our findings promise an advancement of the quantum battery by a general quantum control method.

\section*{Acknowledgments}

We acknowledge grant support from the National Science Foundation of China (Grants No. 11974311 and No. U1801661), Zhejiang Provincial Natural Science Foundation of China (Grant No. LD18A040001), and the Fundamental Research Funds for the Central Universities (Grant No. 2018QNA3004).

\appendix

\section{STA charging performance under various laser shapes}\label{app:other_pulse}

\begin{figure}[htbp]
\centering
\subfigure{
\includegraphics[width=0.45\linewidth]{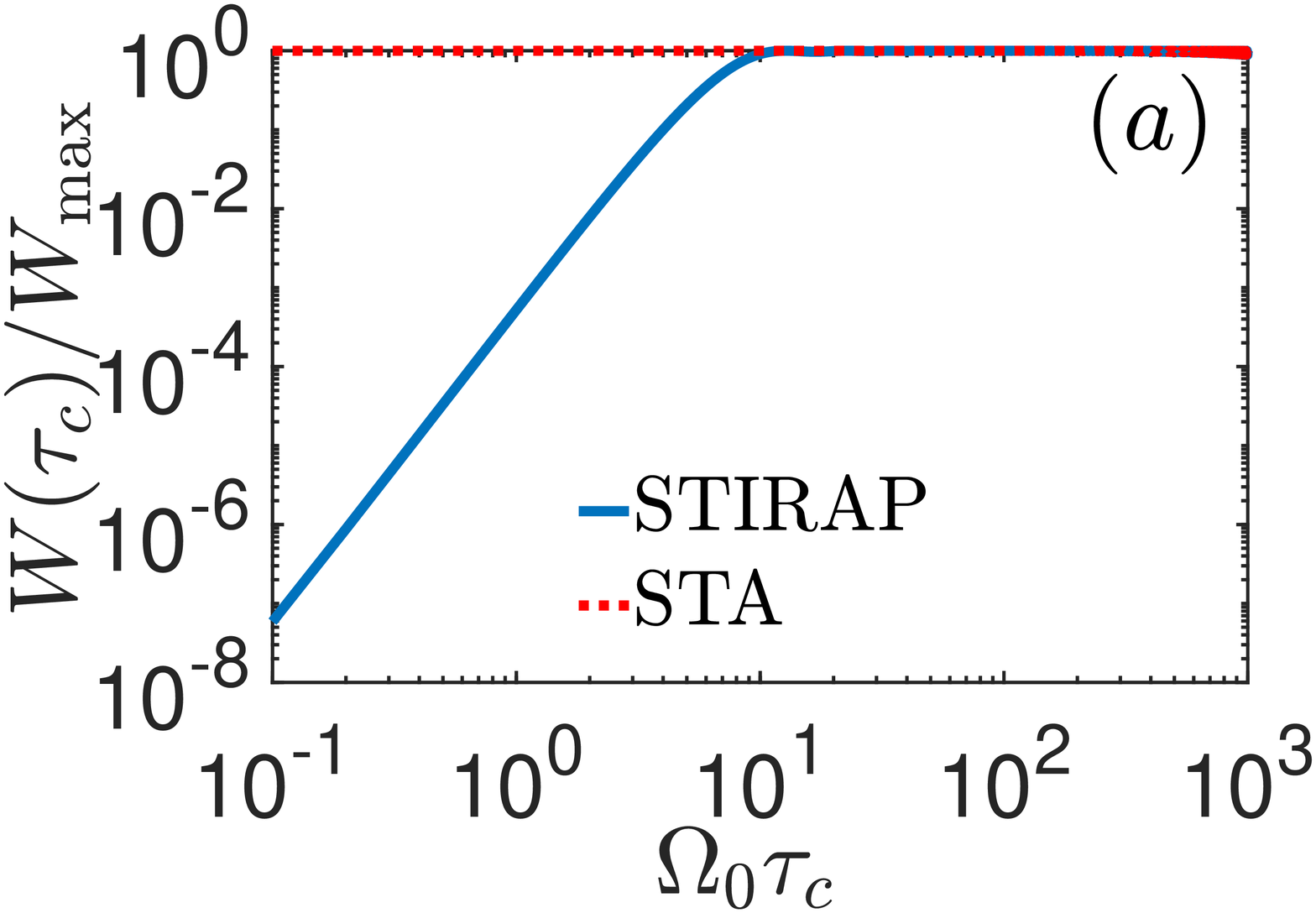}
}
\subfigure{
\includegraphics[width=0.45\linewidth]{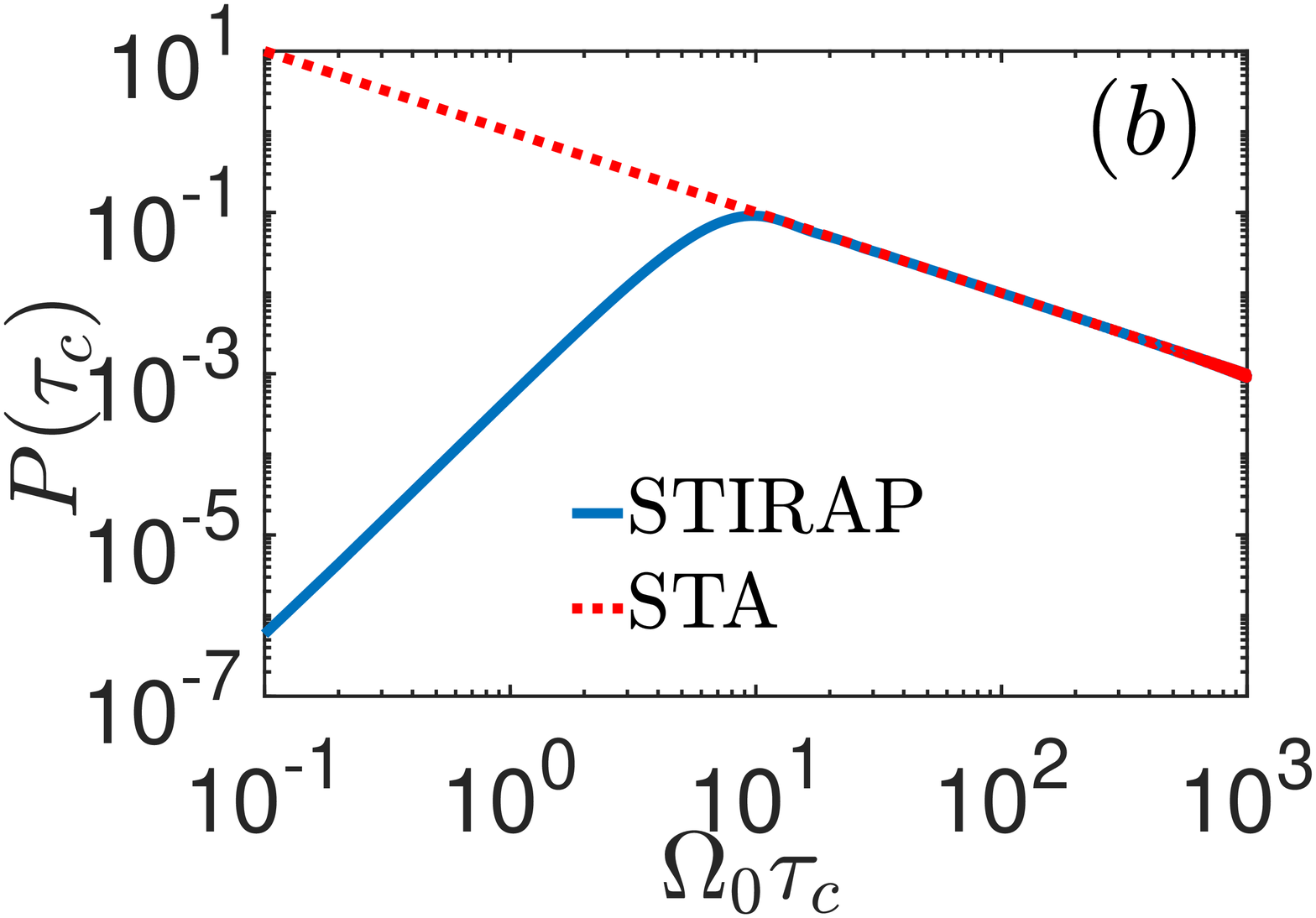}
}
\quad
\subfigure{
\includegraphics[width=0.45\linewidth]{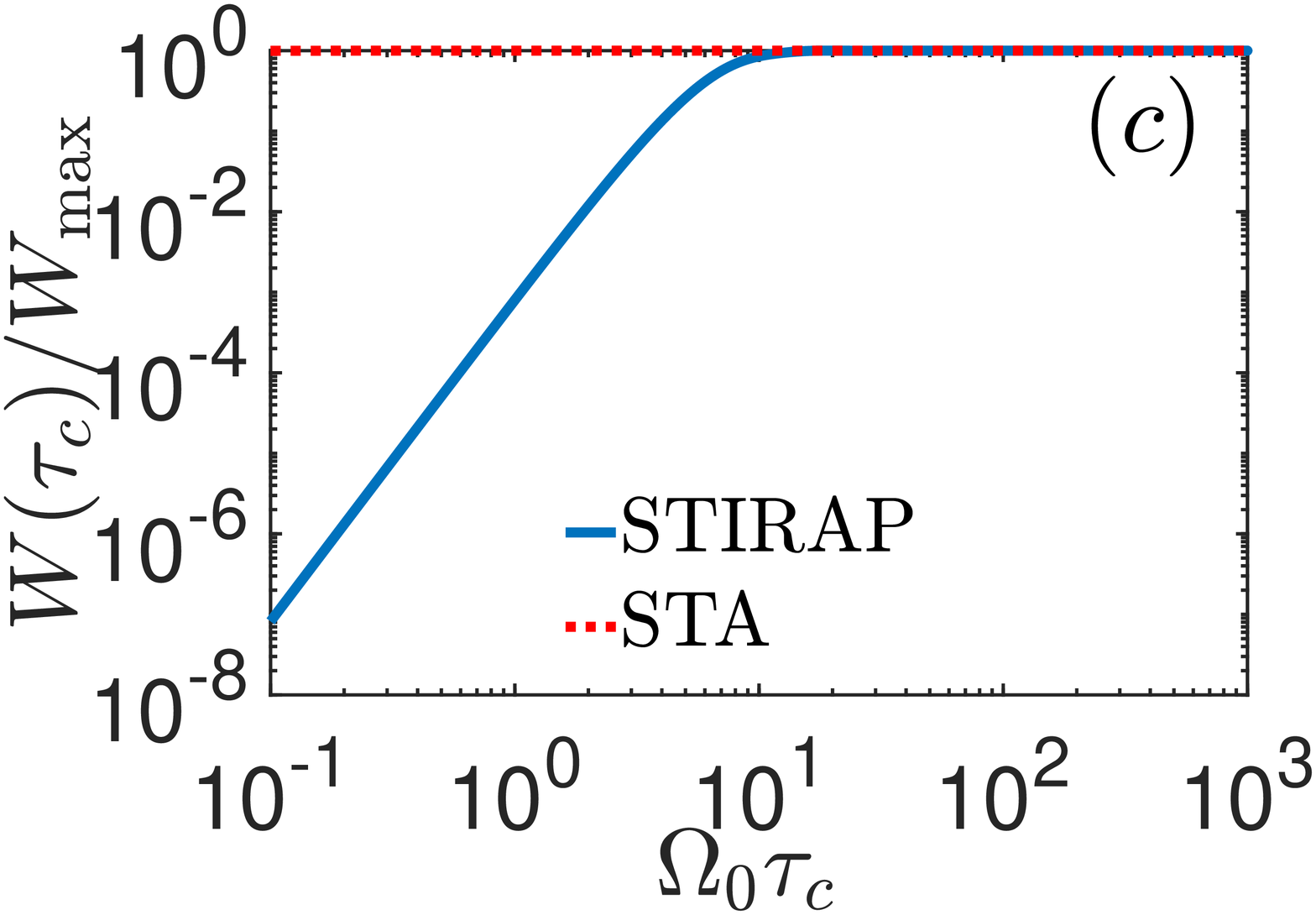}
}
\subfigure{
\includegraphics[width=0.45\linewidth]{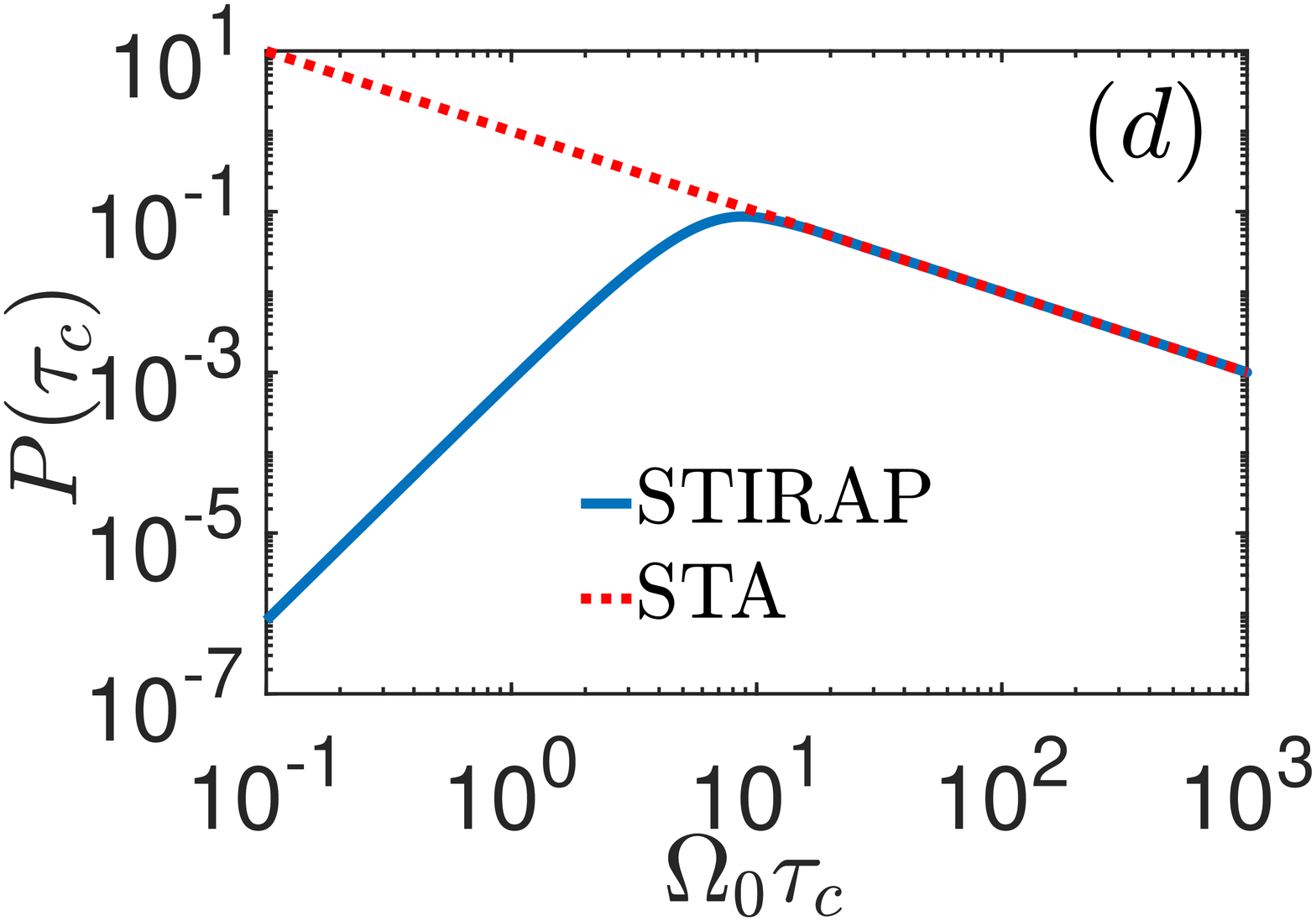}
}
\caption{(Color online) Normalized stored energy and average charge speeds as functions of $\Omega_0\tau_c$ under the stimulated Ramen pulses with the sinusoid shape (a) and (b), and the ramp-like shape (c) and (d). Blue-solid lines and red-dotted lines represent the results for STIRAP and STA charging protocols, respectively.}\label{fig:WP_other_pulse}
\end{figure}

This appendix is contributed to show that the fast and stable charging performance in Sec.~\ref{sec:Gaussian} assisted by STA is irrespective to the shape of the driving pulses. Here we use two extra pulse functions for $\Omega_1(t)$ and $\Omega_2(t)$, and one can analytically obtain $\Omega_a(t)$ through Eq.~(\ref{eq:Oa}).

For the sinusoid pulse,
\begin{equation}\label{eq:sin_O1}
    \Omega_1(t)=\Omega_0\sin^4\left[\frac{\pi(t-\beta)}{\tau_c}\right],
\end{equation}
\begin{equation}\label{eq:sin_O2}
    \Omega_2(t)=\Omega_0\sin^4\left[\frac{\pi(t+\beta)}{\tau_c}\right],
\end{equation}
where $\tau_c$ is the total charging time. The additional CD term reads,
\begin{equation}\label{eq:sin_oa}
    \Omega_a(t)=\frac{\frac{4\pi}{\tau_c}
    \sin\left(\frac{2\pi\beta}{\tau_c}\right)\sin^3\left[\frac{\pi(t-\beta)}{\tau_c}\right]
    \sin^3\left[\frac{\pi(t+\beta)}{\tau_c}\right]}{\sin^8\left[\frac{\pi(t-\beta)}{\tau_c}\right]
    +\sin^8\left[\frac{\pi(t+\beta)}{\tau_c}\right]}.
\end{equation}
In this work, the shape parameters $\beta$ is set to be $\tau_c/10$ for the numerical calculation in Fig.~\ref{fig:WP_other_pulse}(a) and (b).

For the ramp-like pulse,
\begin{equation}\label{eq:ramp_O1}
    \Omega_1(t)=\Omega_0\left[1-\cos\left(\frac{\pi t}{\tau_c}\right)\right],
\end{equation}
\begin{equation} \label{eq:ramp_O2}
    \Omega_2(t)=\Omega_0\cos\left(\frac{\pi t}{\tau_c}\right).
\end{equation}
The additional CD term reads,
\begin{equation} \label{eq:ramp_Oa}
    \Omega_a(t)=\frac{\left(\frac{\pi}{\tau_c}\right)\sin\left(\frac{\pi t}{\tau_c}\right)}{2\cos^2\left(\frac{\pi t}{\tau_c}\right)-2\cos\left(\frac{\pi t}{\tau_c}\right)+1}.
\end{equation}

Similar to the discussion about Fig.~\ref{fig:WP} in Sec.~\ref{sec:Gaussian}, these two pulses are applied to verify the fast and stable charging performance under the STA protocol assisted by CD driving. The normalized stored energy and dimensionless average charging speed are provided in Fig.~\ref{fig:WP_other_pulse}. From either Fig.~\ref{fig:WP_other_pulse}(a) and (b) for the sinusoid pulse or Fig.~\ref{fig:WP_other_pulse}(c) and (d) for the ramp-like pulse, one can find analogical results with the Gaussian pulse as presented in Fig.~\ref{fig:WP}. When the charging time is extremely limited, the stored energy as well as the charing speed of STA is much greater than that of STIRAP by over $7$ or $8$ orders in magnitude. When the charging time is over $\Omega_0\tau=\mathcal{O}(10^1)$, i.e., the adiabatic limit, both $W$ or $P$ of STA and STIRAP are coalescent with each other. Then one can conclude that the STA protocol is irrelevant to the shape of driving pulses, as a fast charging protocol possessing an apparent advantage over the conventional STIRAP.

\nocite{*}
\bibliography{reference}

\end{document}